\documentclass[a4paper, 12pt]{article}
\usepackage{authblk}
\usepackage[top=25truemm,bottom=25truemm,left=20truemm,right=20truemm]{geometry}
\usepackage{setspace}
\usepackage{amsmath,amssymb,mathrsfs,amsbsy,latexsym,amsfonts,amsthm}
\usepackage{fancyhdr}
\usepackage[Symbol]{upgreek}

\usepackage{graphicx}           

\usepackage{color}
\usepackage{slashed}
\usepackage{tensor}
\usepackage{comment}
\usepackage{braket}

\usepackage{ascmac}
\usepackage{here}
\usepackage{bm}

\usepackage{hyperref}

\newcommand{\nn}{\nonumber\\}

\newcommand{\vek}{\vec{k}}

\newcommand{\ba}[1]{\begin{align}#1\end{align}}

\newcommand{\bl}{\big{(}}
\newcommand{\br}{\big{)}}
\newcommand{\bfig}{\begin{figure}[htb!] \centering}
	\newcommand{\efig}{\end{figure}}

\newcommand{\cre}{a^{\dagger}_{\mu}(\vec{k})}

\newcommand{\outst}{\bra{\text{out}}}
\newcommand{\inst}{\ket{\text{in}}}

\makeatletter

\@addtoreset{equation}{section}
\makeatother

\begin{document}
	\setlength{\abovedisplayskip}{15pt}
	\setlength{\belowdisplayskip}{15pt}
\title{
	\bf \Large 
	Conservation Laws from Asymptotic Symmetry  \\ and Subleading Charges in QED	\vskip 0.5cm
}
\author{
	Hayato~Hirai\thanks{\tt hirai@het.phys.sci.osaka-u.ac.jp}
}
\author{
	Sotaro~Sugishita\thanks{\tt sugishita@het.phys.sci.osaka-u.ac.jp}
	\vspace{5mm}
}
\affil{\it\normalsize Department of Physics, Osaka University, Toyonaka, Osaka, 560-0043, Japan}
\setcounter{Maxaffil}{0}
\date{}

\maketitle
\thispagestyle{fancy}
\renewcommand{\headrulewidth}{0pt}

\begin{abstract}
We present several results on memory effects, asymptotic symmetry and soft theorems in massive QED. We first clarify in what sense the memory effects are interpreted as the charge conservation of the large gauge transformations, and derive the leading and subleading memory effects in classical electromagnetism. We also show that the sub-subleading charges are not conserved without including contributions from the spacelike infinity. Next, we study QED in the BRST formalism and show that parts of large gauge transformations are physical symmetries by justifying that they are not gauge redundancies. Finally, we obtain the expression of charges associated with the subleading soft photon theorem in massive scalar QED. 
\end{abstract}

\newpage
\setcounter{tocdepth}{2}
\tableofcontents

\newpage
\section{Introduction and Summary}\label{sec:intro}

Asymptotic symmetries in gauge theories and also gravity have been investigated with their relations to soft theorems and memory effects (see \textit{e.g.}, \cite{Strominger:2017zoo} for a review).\footnote{Similar relations were also investigated for massless scalar theories \cite{Campiglia:2017dpg, Hamada:2017atr, Campiglia:2017xkp}. In particular, it was shown \cite{Hamada:2017atr} that there is a memory effect of pion or axion radiation associated with the soft pion theorem.}  

In quantum electrodynamics (QED), the soft photon theorem \cite{Yennie:1961ad, Weinberg:1965nx} is a universal relation between a scattering amplitude with a soft photon and one without the photon. 
In \cite{He:2014cra,Campiglia:2015qka,Kapec:2015ena}, it was shown that the soft photon theorem is equivalent to the Ward-Takahashi identities for the asymptotic symmetry, 
which is an infinite dimensional subgroup of $U(1)$ gauge transformations taking nonzero finite values at far distant regions. 
In addition, it has been known that the subleading term in the soft expansion of the scattering amplitude is also universal \cite{Low:1954kd,Low:1958sn,Burnett:1967km,GellMann:1954kc}. 
In \cite{Lysov:2014csa, Campiglia:2016hvg, Conde:2016csj}, it was argued that, as well as the leading soft theorem, the subleading photon theorem can be interpreted as the Ward-Takahashi identity of an infinite number of symmetries for QED with massless charges. 

On the other hand, the memory effect is a phenomena for the classical radiation of massless fields. 
It was first pointed out in gravitational theory \cite{ZelPol} (see also \cite{Braginsky1987,Christodoulou:1991cr,Thorne:1992sdb}). 
If there is an event producing gravitational radiation with finite duration, it causes a shift of the metric perturbation from far past to far future at the far distant region of the event. 
It is not surprise that we also have the similar phenomena in classical electromagnetism \cite{Bieri:2013hqa, Tolish:2014bka, Susskind:2015hpa} because the memory effects essentially follow from the properties of the retarded Green's function of massless particles. 
It has been known \cite{Yennie:1961ad} that the soft factor in the leading soft photon theorem appears in the solution of the classical radiation induced by the scattering of point charges. In fact, the leading and subleading soft factors are related with leading and subleading memories \cite{Strominger:2014pwa, Pasterski:2015tva, Mao:2017wvx, Hamada:2018cjj}.

However, to our knowledge, there has been no derivation of the memory effects as a direct consequence of the asymptotic symmetries, although it was pointed out that the shifts of radiation fields can be realized as transformations of the asymptotic symmetries \cite{Strominger:2014pwa}. 
In literatures, memory effects are derived by solving the equations of motion \cite{ZelPol, Bieri:2013hqa, Tolish:2014bka, Strominger:2014pwa, Pasterski:2015tva, Susskind:2015hpa, Mao:2017axa, Mao:2017wvx, Hamada:2018cjj} or from the soft theorems \cite{Strominger:2014pwa, Pasterski:2015tva}. 

In section~\ref{sec:charge conservation} of this paper, we will explicitly illustrate in classical electromagnetism that the electromagnetic memory effect is derived from the conservation law of the large gauge symmetry. We will work in the Lorenz gauge $\partial_\mu A^\mu=0$. Then the large gauge symmetry is the residual gauge transformations $A_\mu \to A_\mu + \partial_\mu \epsilon$ with parameters $\epsilon(x)$ which satisfy $\Box \epsilon(x)=0$ to  keep the gauge fixing condition and approach angle-dependent nontrivial functions $\epsilon^{(0)}(\Omega)$ at infinity $r\to \infty$.
Since $\epsilon^{(0)}(\Omega)$ is an arbitrary function on two-sphere, the symmetry is infinite-dimensional. 
The memory effect will be derived from the current conservation associated with the symmetry. 
When we investigate the asymptotic behaviors of fields, we should be careful about the treatment of asymptotic infinities. 
In particular, the timelike infinity $i^\pm$ na\"ively looks collapsed to a point (\textit{i.e.}, two-sphere) in the Penrose diagram, although it is actually an infinite-time limit of a three-dimensional constant time-slice. 
Thus, we first consider the gauge-charge conservation in a specific finite region (see Fig.~\ref{fig:region}) and then take a limit such that the boundaries of the region approach asymptotic infinities of Minkowski space.
Due to the specific choice of region, we can treat the contributions from timelike infinities $i^\pm$ and null infinities $\mathscr{I}^\pm$ separately. 
The region also has a boundary approaching to the spacelike infinity $i^0$. 
Nevertheless, it will be shown that the contributions from $i^0$ vanish, and this fact leads to the asymptotic charge conservation laws which turn out to be the memory effect [see \eqref{lead_memory} and \eqref{lead_charge_consv}]. 
The resulting formula \eqref{lead_memory} shows how the radiation field shifts when there is a nontrivial change of the distribution of charged matters from far past to far future. 

The choice of the finite region also enables us to evaluate the subleading terms in the gauge-charge conservation in a safe manner. 
Through the evaluations in subsection~\ref{subsec:sub_memory}, the subleading memory effect will be shown [see \eqref{eq:sub_soft_cl} and \eqref{sub_charge_consv}]. It will also be shown that there is no sub-subleading memory effect because at this order we should take account of the contributions to the gauge-charge from the spacelike infinity $i^0$. 
In appendix~\ref{sec:app_a}, the leading and subleading memory effects are confirmed for a concrete setup. 

It is also not obvious whether the large gauge symmetry is physical symmetry, because it na\"ively seems to be a part of gauge redundancies. 
One approach to this question is the canonical treatment of radiation data as in \cite{He:2014laa, He:2014cra} (see also \cite{Ashtekar:1981sf,Frolov:1977bp}). 
In section~\ref{sec:brst}, we will address this question from a different perspective. 
We will consider QED in the BRST formalism \cite{Becchi:1975nq, Tyutin:1975qk}, and argue that the large gauge transformations should be regarded as physical symmetry. In fact if we regard the large gauge transformations as gauge redundancies, the physical scattering data which we can treat is too restricted. We will also see that the large gauge transformations automatically become physical transformations of the Cauchy data if ghost fields can be expanded in the standard Fourier modes.  
In subsection~\ref{subsec:leadingsoft}, we will review for completeness that the leading soft theorem is equivalent to the Ward-Takahashi identity for the asymptotic symmetry in our notation. This review may be useful to read section~\ref{sec:subleading}.

In section~\ref{sec:subleading}, we will interpret the subleading soft photon theorem as the Ward-Takahashi identity for the asymptotic symmetry.
As mentioned above, such interpretation was given in \cite{Lysov:2014csa, Campiglia:2016hvg, Conde:2016csj} for massless QED. However, massive particles were not included in their analysis because a different treatment are needed for massive particles. We will carry out this treatment for massive charged scalars, and obtain the expression of the hard charge operators in massive scalar QED. 
Asymptotic behaviors of the massive scalar are summarized in appendix~\ref{app:asympt_hard}. 
Some complicated calculations in the derivation of the hard charge operators are confined in appendix~\ref{cal}. 

Finally some discussions will be given in section~\ref{sec:disc}.

\section{Infinite number of conserved charges}\label{sec:charge conservation}
In this section, we illustrate the existence of an infinite number of asymptotically conserved charges associated with large gauge transformation in the classical electromagnetism. 
We represent the matter current for massive charged particles by $j_{mat}^\mu$, which is the source in Maxwell's equation $\partial_\nu F^{\nu\mu}=-j_{mat}^\mu$. Here, we assume that the charged particles behave as free particles except for a small region where they scatter, and we ignore the back-reaction. We also impose the initial condition that there is no radiation before the scattering of charged particles.

\begin{figure}[htbp]
	\vspace{0.5cm}
	\begin{center}
		\includegraphics[width=5.0cm]{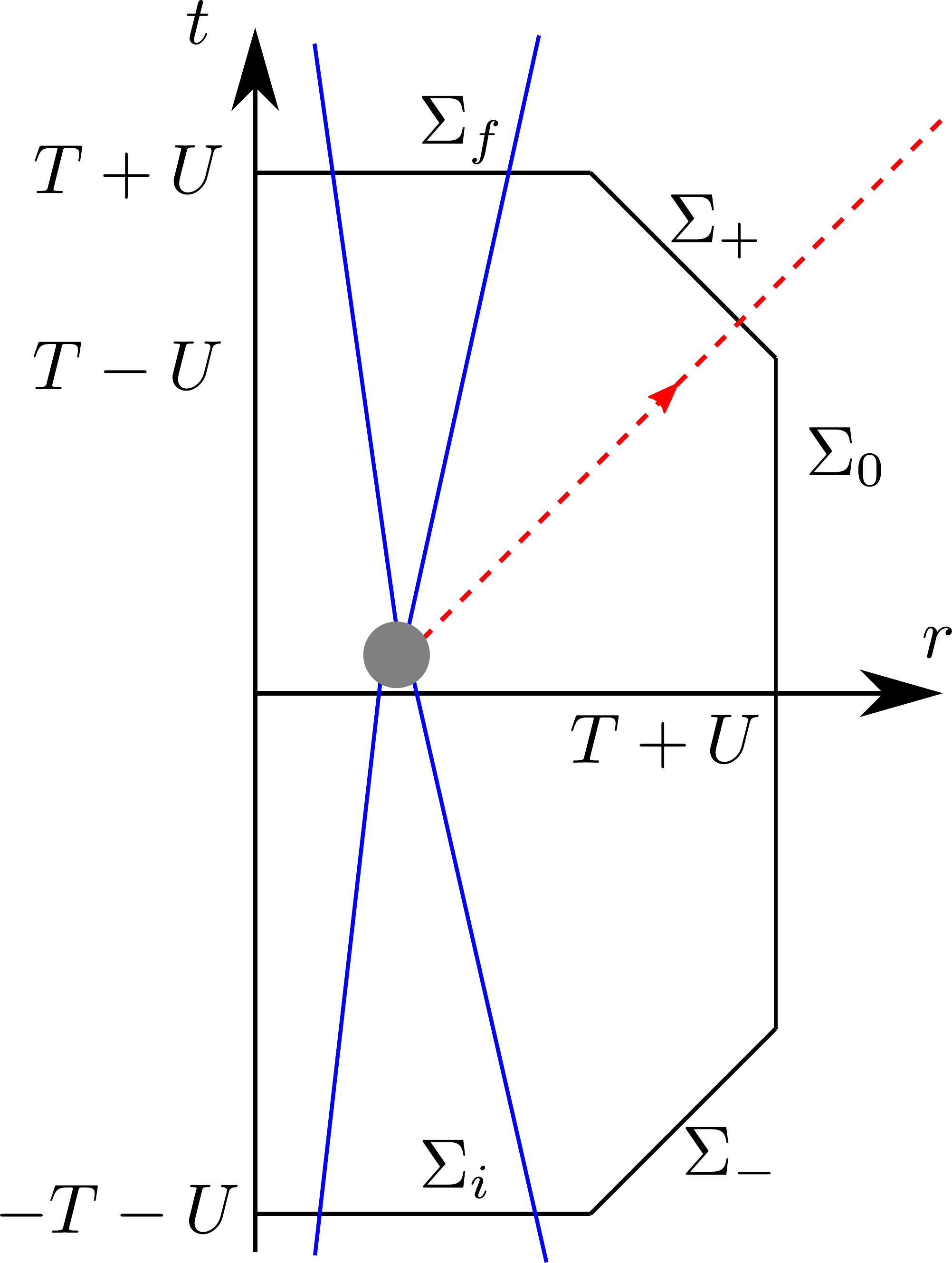}
		\caption{The region where we consider the current conservation. The directions along the two sphere $S^2$ are suppressed in this figure. The blue lines represent  trajectories of massive charged particles, which scatter at a small region. The red line represents a direction of the radiation emitted from this scattering. The region is parametrized by two parameters $T$ and $U$. The parameter $T$ is so large that all of the given massive particles go through the surface $\Sigma_i$ and $\Sigma_f$, and $U$ is also so large that any radiation coming from the scattering region passes through $\Sigma_+$.  
		}
		\label{fig:region}
	\end{center}
\vspace{0.5cm}
\end{figure}

The conserved current for the gauge transformation with gauge parameter $\epsilon(x)$ is given by
\begin{align}
\label{class_current}
J^\mu = F^{\mu\nu}\partial_\nu \epsilon +j_{mat}^\mu \epsilon. 
\end{align}
We now consider the integration of the current conservation equation $\partial_\mu J^\mu=0$ over the region represented in Fig.~\ref{fig:region}. 
The region is parametrized by two parameters $T$ and $U$, 
and has five boundaries, $\Sigma_f, \Sigma_i, \Sigma_+, \Sigma_-, \Sigma_0$. 
The $\Sigma_f$ and $\Sigma_i$ are time-slices at far future and past $t=\pm(T+U)$ with $0\leq r \leq T-U$. $\Sigma_0$ is a timelike surface $r=\text{const.}=T+U$ with $-T+U<t<T-U$, respectively. 
$\Sigma_{\pm}$ are null surfaces where $\pm t+r=\text{const.}=2T$. 
We take the parameter $T$ is so large such that the massive matter current $j_{mat}^\mu$ vanishes on $\Sigma_\pm$ and $\Sigma_0$. 
We also take $U$  so large  that any electromagnetic radiation coming from the scattering region  goes through the null surface $\Sigma_+$. Later, we take the limit first $T\to \infty$ and then $U\to \infty$ such that $\Sigma_\pm$ become the future and past null infinities $\mathscr{I}^\pm$, where $\mathscr{I}^+$ ($\mathscr{I}^-$) is parametrized by the retarded (advanced) time $u= t-r$ ($v=t+r$) and angular coordinates $\Omega^A$.
The ordering of the limits, $T\to \infty$ before $U\to \infty$, is crucial for our derivation of the memory effect formulae. 
The integration of the current conservation proves that the sum of the surface integral of the current over each boundary vanishes: 
\begin{align}
0=\int d V\, \partial_\mu J^\mu = Q_f + Q_+ +Q_0 - Q_- - Q_i
\label{total_cons}
\end{align}
with 
\begin{align}
Q_a \equiv \int_{\Sigma_a} d S^\mu J_\mu\,, \quad (a=f,i,0,+,-) \,,
\end{align}
where we choose the surface element $dS^\mu$ is future-directed when the surface is spacelike or null.

We consider a residual large gauge parameter $\epsilon(x)$ in the Lorenz gauge, which is a solution of  $\Box \epsilon(x)=0$ asymptotically approaching an arbitrary function $\epsilon^{(0)}(\Omega)$ on the two-sphere,\footnote{We represent the coordinates on the unit two-sphere by $\Omega^A$ (A=1,2) with metric $\gamma_{AB}$. } independently of $u$, on $\mathscr{I}^+$. The solution is given by \cite{Campiglia:2015qka, Strominger:2017zoo}
\begin{align}
&\epsilon(x)=\int d^2 \Omega' \sqrt{\gamma(\Omega')}\,  G(x;\Omega') \epsilon^{(0)}(\Omega')\,,\label{LGP}\\
& G(x;\Omega') = -\frac{1}{4\pi} \frac{x^\mu x_\mu}{(q^\mu x_\mu )^2} \quad \text{with} \quad q^\mu=(1,\hat{q}(\Omega')) 
\label{green_p}, 
\end{align}   
where $\hat{q}(\Omega)$ is a three-dimensional unit vector parametrized by $\Omega^{A}$.\footnote{More precisely, Green's function is defined as 
	$G(x;\tilde{\Omega}) =-\frac{1}{8\pi} \lim_{\varepsilon\to 0}\left[\frac{x^\mu x_\mu}{(q^\mu x_\mu -i \varepsilon)^2}+\frac{x^\mu x_\mu}{(q^\mu x_\mu +i \varepsilon)^2}\right]$.}
One can find that this $\epsilon(x)$ satisfies the antipodal matching condition, \textit{i.e.}, $\epsilon(t, r, \Omega)$ approaches $\epsilon^{(0)}(\bar{\Omega})$ on $\mathscr{I}^-$ where $\bar{\Omega}^{A}$ denotes the antipodal point of $\Omega^A$. 
The current conservation \eqref{total_cons} for large gauge transformations characterized above $\epsilon(x)$ will lead to the memory effects. 

\subsection{Leading memory effect}\label{LMF}
We now see that $Q_0$, which is defined on $\Sigma_0$ as 
\begin{align}
Q_0= \int^{T-U}_{-T+U} \!\!\! dt\, d^2 \Omega \sqrt{\gamma} r^2 J^r |_{r=T+U}\,,  
\end{align}
vanishes in the limit $T\to \infty$ without the limit $U\to \infty$. 
Noting that the current can be represented as a total derivative $J^r= \partial_\mu (F^{r\mu} \epsilon)$, 
$Q_0$ splits into the integrations over two spheres on the future and past boundaries of $\Sigma_0$ as 
\begin{align}
\label{Q0_sum}
Q_0=\int d^2 \Omega \sqrt{\gamma} r^2 F_{tr} \epsilon |_{t= T-U, r=T+U} - \int d^2 \Omega \sqrt{\gamma} r^2 F_{tr} \epsilon |_{t=- T+U, r=T+U}\,.
\end{align}
Since any radiation cannot reach $\Sigma_0$ due to our choice of the region, 
the field strength $F_{tr}$ is the Coulombic electric field created by free-moving charged particles before the scattering, and it behaves as $F_{tr} \sim \mathcal{O}(T^{-2})$ in the large $T$ limit. 
The leading $\mathcal{O}(T^{-2})$ terms of the Coulombic electric fields are independent of $U$, which are represented as  
\begin{align}
&\lim_{T\to \infty} r^2 F_{tr}(t,r,\Omega) |_{t= T-U, r=T+U} \equiv F^{+(2)}_{tr}(\Omega)\,,  \label{F+2}\\
&\lim_{T\to \infty} r^2 F_{tr}(t,r,\Omega) |_{t= -T+U, r=T+U} \equiv F^{-(2)}_{tr}(\Omega)\,,  \label{F-2}
\end{align}
and they satisfy the antipodal matching condition (see appendix~\ref{app_a})
\begin{align}
\label{em_match}
F^{+(2)}_{tr}(\Omega)= F^{-(2)}_{tr}(\bar\Omega)\,.  
\end{align} 
We thus obtain  
\begin{align}
\lim_{T\to \infty} Q_0 = \int d^2 \Omega \sqrt{\gamma} F^{+(2)}_{tr}(\Omega) \epsilon^{(0)} (\Omega) - \int d^2 \Omega \sqrt{\gamma} F^{-(2)}_{tr}(\Omega) \epsilon^{(0)} (\bar\Omega)  
\end{align}
where we used the fact that $\epsilon(x)$ approaches $\epsilon^{(0)}(\Omega)$ near $\mathscr{I}^+$ and $\epsilon^{(0)}(\bar\Omega)$ near $\mathscr{I}^-$. 
Therefore, due to the matching condition \eqref{em_match}, 
$Q_0$ vanishes in the limit $T\to \infty$:  
\begin{align}
\lim_{T\to \infty} Q_0=\int d^2 \Omega \sqrt{\gamma} [F^{+(2)}_{tr}(\Omega)-F^{-(2)}_{tr}(\bar\Omega)] \epsilon^{(0)} (\Omega) = 0\,. 
\end{align}

Thus, in the limit $T\to \infty$, the conservation equation \eqref{total_cons} indicates 
\begin{align}
\label{cc_t_infty}
\lim_{T\to \infty }(Q_f + Q_+) = \lim_{T\to \infty }(Q_i + Q_-)\,. 
\end{align}
The $Q_f$ and $Q_i$ are given by  
\begin{align}
Q_f &= \int^{T-U}_0 \!\!\! dr d^2 \Omega \sqrt{\gamma}r^2  (F^{t i} \partial_i \epsilon + j_{mat}^t \epsilon)|_{t=T+U}, \\
Q_i &= \int^{T-U}_0 \!\!\! dr d^2 \Omega \sqrt{\gamma}r^2  (F^{t i} \partial_i \epsilon + j_{mat}^t \epsilon)|_{t=-T-U}.
\end{align}
The field strength $F^{t i}$ in the above integrands is the Coulombic electric field created by free-moving charged particles since the radiation does not reach $\Sigma_f$ and $\Sigma_i$. Thus, $Q_f$ and $Q_i$ are the usual gauge charges for free-moving charged particles and their Coulomb-like potential in the three-dimensional ball with radius $T-U$ at time $t=\pm (T+U)$. In particular, if $\epsilon$ is a constant, $Q_f$ and $Q_i$ equal to the total electric charges ($\times \epsilon$). Eq.~\eqref{cc_t_infty} just means that such gauge charges do not conserve for the nontrivial large gauge parameters $\epsilon(x)$ unless we include the contributions from the radiation $Q_\pm$.\footnote{A similar statement was given for the case without massive charged fields in \cite{Hamada:2017bgi}.}
It should be remarked that the conservation laws \eqref{cc_t_infty} holds only asymptotically, \textit{i.e.}, in the limit $T \to \infty$. 
We call this kind of conservation ``asymptotic conservation". 
Since we can take arbitrary functions $\epsilon^{(0)}(\Omega)$ on two-sphere, we have an infinite number of asymptotic conservation laws. 

Using the retarded time $u=t-r$,\footnote{The line element in the retarded coordinates takes the form
	$ds^{2}=-du^{2}-2dudr+r^{2}\gamma_{AB}d\Omega^{A}d\Omega^{B}$.} 
$Q_+$ is written as 
\begin{align}
\label{def_Q+}
Q_+= \int^{2U}_{-2U}\!\!du d^2 \Omega \sqrt{\gamma} r^2 (F^{ru}\partial_u \epsilon + F^{r A}\partial_A \epsilon)|_{r=T-u/2} \,. 
\end{align} 
In the large $T$ limit, the integration region approaches to the subregion $-2U\leq u \leq 2U$ in $\mathscr{I}^+$. Near $\mathscr{I}^+$, $A_u$ and $A_r$ decay faster, and only the angular components $A_B$ behave as $\mathcal{O}(1)$. We represent the $\mathcal{O}(1)$ components  by $A^{(0)}_B(u,\Omega)$.  Then, we obtain\footnote{We have $F^{rB}=-\partial_u A^B +\partial_r A^B +\frac{\gamma^{BC}}{r^2}\partial_C(A_u-A_r)$.} 
\begin{align}
\lim_{T\to \infty}Q_+ &=  -\int^{2U}_{-2U}\!\!du d^2 \Omega \sqrt{\gamma} \gamma^{AB}\partial_u A^{(0)}_B \partial_A \epsilon^{(0)} \nn
&=-\int d^2 \Omega \sqrt{\gamma} \gamma^{AB}[ A^{(0)}_B(u=2U)-A^{(0)}_B(u=-2U)] \partial_A \epsilon^{(0)}\,.
\label{finite_soft}
\end{align}
On the other hand, for our setup or our initial condition, there is no radiation at $\Sigma_-$, and thus $\lim_{T\to \infty} Q_-=0$. 
Therefore, we have the following memory effect formula:  
\begin{align}
\int d^2 \Omega \sqrt{\gamma} \gamma^{AB}[ A^{(0)}_B(u=2U)-A^{(0)}_B(u=-2U)] \partial_A \epsilon^{(0)}= \lim_{T\to \infty} (Q_f-Q_i)\,. 
\label{lead_memory}
\end{align}
This equation holds for any function $\epsilon^{(0)}(\Omega)$ on the two-sphere. 
It implies that the shift of $A^{(0)}_B$ on $\mathscr{I}^+$ is related to the change of the gauge charges between $Q_f$ and $Q_i$.\footnote{The $\Omega$-independent part of the shift cannot be determined from this formula.}

If we take the limit $U \to \infty$, the charge from the radiation \eqref{finite_soft} becomes the so called `soft charge' \cite{He:2014cra, Campiglia:2015qka}
\begin{align}
\label{cl_lead_soft_charge}
Q^{\text{lead},+}_S \equiv -\int_{\mathscr{I}^+}\!\!du d^2 \Omega \sqrt{\gamma} \gamma^{AB}\partial_u A^{(0)}_B \partial_A \epsilon^{(0)}. 
\end{align}
Although $\lim_{T\to \infty}Q_-$ vanishes in our setup, if we consider the general situation that  electromagnetic radiation exists initially, $\lim_{U\to \infty}\lim_{T\to \infty}Q_-$ becomes the soft charge $Q^{\text{lead},-}_S$ defined on $\mathscr{I}^-$. $Q_f$ and $Q_i$ become the so called hard charges $Q^{\text{lead},\pm}_H$ in the limit. Note that the hard charges include not only matter currents but also the contributions from the Coulombic electric field produced by the charged matters. 

In the limit $U \to \infty$, we thus obtain 
\begin{align}
Q^{\text{lead},+}_S+Q^{\text{lead},+}_H=Q^{\text{lead},-}_S+Q^{\text{lead},-}_H\,.
\label{lead_charge_consv}
\end{align}
We note again that the conservation laws \eqref{lead_charge_consv} come from 
the fact that the charge on spacelike infinity, $\lim_{T\to \infty} Q_0$, vanishes due to the antipodal matching condition.  
In other words, the total asymptotic charge $Q^{\text{lead},+}_S+Q^{\text{lead},+}_H$ equals to the integral on two-sphere at the past boundary of $\mathscr{I}^+$ at $u=-\infty$ as
\begin{align}
\label{eq:tot_+}
Q^{\text{lead},+}_S+Q^{\text{lead},+}_H = -\int_{\mathscr{I}^+_-} d^2 \Omega \sqrt{\gamma} F^{+(2)}_{tr}(\Omega) \epsilon^{(0)} (\Omega)\,,
\end{align}
because the current $J^\mu$ is a total derivative, 
and $Q^{\text{lead},-}_S+Q^{\text{lead},-}_H$ also equals to the integral on two-sphere at the future boundary of $\mathscr{I}^-$ as
\begin{align}
\label{eq:tot_-}
Q^{\text{lead},-}_S+Q^{\text{lead},-}_H = -\int_{\mathscr{I}^-_+} d^2 \Omega \sqrt{\gamma} F^{+(2)}_{tr}(\Omega) \epsilon^{(0)} (\bar{\Omega})\,,
\end{align}
which is equal to \eqref{eq:tot_+}. 
We also remark that the surface $\Sigma_\pm\cup \Sigma_{f/i}$ becomes the Cauchy surface in the limit $U\to \infty$ after $T\to \infty$. 
Therefore, $Q^{\text{lead},\pm}_S+Q^{\text{lead},\pm}_H$ is actually a charge defined on the asymptotic Cauchy surface.

\subsection{Subleading memory effect}\label{subsec:sub_memory}
We have seen that the current conservation equation \eqref{total_cons} leads to the formula of the leading memory effect \eqref{lead_memory} in the large $T$ limit because the charge from spacelike infinity, $Q_0$, vanishes in the limit due to the antipodal matching. 
The subleading memory effect can also be obtained by considering the corrections in  large $T$ expansion.  

We first consider the correction to $Q_+$ defined as eq.~\eqref{def_Q+}. 
On $\Sigma_+$, the gauge parameter $\epsilon(x)$ in the large $T$ limit is expanded using the formula \eqref{large-r_ep} as 
\begin{align}
\label{epsilon_exp}
\epsilon(u, r=T-u/2, \Omega) = \epsilon^{(0)}(\Omega)- \frac{u \log{\frac{|u|}{2 T}}}{2 T} \Delta_{\mathrm{S}^2}\epsilon^{(0)}(\Omega) + \mathcal{O}(T^{-1})\,, 
\end{align}
where $\Delta_{\mathrm{S}^2}$ is the Laplacian on the unit two-sphere. Note that the correction to $\epsilon^{(0)}(\Omega)$ starts from $\mathcal{O}(T^{-1}\log{T})$. 
In appendix \ref{app:asympt_rad}, we give general expansions of radiation fields which are compatible with this large gauge parameters. 
Using the expansions \eqref{epsilon_exp}, \eqref{eqapp_au}, \eqref{eqapp_ar} and \eqref{eqapp_ab}, we find that 
the first correction to $Q_+$ is $\mathcal{O}(T^{-1}\log T)$, and $Q_+$ takes the form 
\begin{align}
Q_+ &=-\int^{2U}_{-2U}\!\!du d^2 \Omega \sqrt{\gamma} \partial_u A^{(0)}_B \gamma^{BA}\partial_A\epsilon^{(0)}
-(Q_+^\text{log}+Q_+^{\text{log}\prime}) \frac{\log{T}}{T} + \mathcal{O}(T^{-1})\,,
\end{align}
where the first term is just the leading soft charge \eqref{finite_soft}, and the second term is the first correction. 
Here, $Q_+^\text{log}$ and $Q_+^{\text{log}\prime}$ are given by
\begin{align}
\label{q+log}
Q_+^\text{log} &= \frac12 \int^{2U}_{-2U}\!\!du d^2 \Omega \sqrt{\gamma} u \partial_u A^{(0)}_B \gamma^{BA}\partial_A \Delta_{\mathrm{S}^2} \epsilon^{(0)}\nn
&=-\frac12 \int^{2U}_{-2U}\!\!du d^2 \Omega \sqrt{\gamma}  \epsilon^{(0)} u \partial_u \Delta_{\mathrm{S}^2}  \nabla^B A^{(0)}_B 
\,,
\\
\label{q+logpr}
Q_+^{\text{log}\prime} &= 
-\frac12 \int^{2U}_{-2U}\!\!du d^2 \Omega \sqrt{\gamma} \left[\left(
A_r^{(1)}+   \nabla^B A^{(0)}_B  + 2 C_u^{(1)}
\right)\Delta_{\mathrm{S}^2} \epsilon^{(0)}
+2 \gamma^{AB}
\left(\partial_u C_A^{(1)} - \partial_A C_u^{(1)}
\right)\partial_B \epsilon^{(0)}\right]
\,,
\end{align}
where $\nabla_B$ denotes the covariant derivative on the unit two-sphere w.r.t. the metric $\gamma_{AB}$, and the derivative with upper index is defined as $\nabla^A\equiv \gamma^{AB}\nabla_{B}$.
The definition of $C_u^{(1)}$, $A_r^{(1)}$ and  $C_A^{(1)}$ are given in \eqref{eqapp_au}, \eqref{eqapp_ar} and \eqref{eqapp_ab}.

On spatially distant surface $\Sigma_0$, the leading part of the charge $Q_0$ in large $T$ expansion is $\mathcal{O}(T^{-1})$ as shown in eq.~\eqref{Q_0_app} in appendix \ref{app_a}, and it does not have any $\mathcal{O}(T^{-1}\log T)$ term.  
Thus, the coefficient of $T^{-1}\log{T}$ is also conserved without contribution from $\Sigma_0$ like the leading memory. 
The contributions from future and past timelike infinities are extracted as 
\begin{align}
Q_{f,i}^\text{log} = \lim_{T \to \infty} \frac{T^2}{\log T} \frac{d Q_{f,i}}{d T}. 
\end{align}
Using these symbols, the finite $U$ version of the subleading memory effect formula takes the form 
\begin{align}
-\frac12 \int^{2U}_{-2U}\!\!du d^2 \Omega \sqrt{\gamma}  \epsilon^{(0)} u \partial_u \Delta_{\mathrm{S}^2} \nabla^B A^{(0)}_B   =
-Q_+^{\text{log}\prime}
-Q_{f}^\text{log} +Q_{i}^\text{log}\,.
\label{sublead_memory}
\end{align}

In the limit $U\to \infty$, the subleading radiation charge $Q_+^\text{log}$ becomes 
\begin{align}
\label{eq:sub_soft_cl}
Q^{\text{sub},+}_S \equiv \lim_{U \to \infty} Q_+^\text{log} = -\frac12 \int_{\mathscr{I}^+}\!\!\!du d^2 \Omega \sqrt{\gamma}  \epsilon^{(0)} u \partial_u \Delta_{\mathrm{S}^2}  \nabla^B A^{(0)}_B,
\end{align}
which agrees, up to a numerical factor, with the electric-type subleading soft charge in \cite{Campiglia:2016hvg}, and it is shown that the charge also agrees with that in \cite{Lysov:2014csa} by taking their vector field $Y^A$ on unit two-sphere as $Y^A \propto \nabla^A \epsilon^{(0)}$. 
We also represent the other contributions in the limit $U\to \infty$ including the initial radiation on $\Sigma_-$ by\footnote{
	As shown in appendix \ref{app_comp_memo}, $Q_+^{\text{log}\prime}$ generally diverges in the limit $U\to \infty$, and $Q_f^\text{log}$ also does. 
	Nevertheless, the combination $(Q_+^{\text{log}\prime}+Q_f^\text{log})$ is finite as far as we know. 
	}
\begin{align}
Q^{\text{sub},-}_S \equiv \lim_{U \to \infty} Q_-^\text{log}\,,\quad
Q^{\text{sub},+}_H \equiv \lim_{U \to \infty}
(Q_+^{\text{log}\prime}+Q_f^\text{log})\,,\quad
Q^{\text{sub},-}_H \equiv \lim_{U \to \infty} 
(Q_-^{\text{log}\prime}+Q_i^\text{log})\,.
\end{align}
We then obtain the subleading charge conservation 
\begin{align}
Q^{\text{sub},+}_S+Q^{\text{sub},+}_H=Q^{\text{sub},-}_S+Q^{\text{sub},-}_H\,.
\label{sub_charge_consv}
\end{align}

Before closing this section, we briefly comment on the sub-subleading order. 
We have seen that the charge $Q_0$ on spacelike infinity is $\mathcal{O}(T^{-1})$, which is the same order as the sub-subleading corrections to $Q_a$ $(a=+,-,f,i)$.  
Therefore, we conclude that there is no asymptotic conservation at the sub-subleading order without including the contribution from $Q_0$ (see also similar discussions in \cite{Campiglia:2016hvg, Conde:2016csj}).\footnote{The arguments that there is no subsubleading charge were given in \cite{Campiglia:2016hvg, Conde:2016csj}. However, the reason seems to be different from ours. Their reason is that the sub-subleading charge has an inevitable divergence. On the other hand, from our construction, the charges for the large gauge transformations are ``conserved" if we take account of the contributions from spacelike infinity. } 
This conclusion is probably related to the fact that there is no sub-subleading soft photon theorem in QED (see \cite{Hamada:2018vrw} for the discussion that the soft expansion of amplitudes is not associated with a universal soft factor at the sub-subleading order).


\section{Asymptotic symmetry as physical symmetry in QED}
\label{sec:brst}
We have seen that the current conservation for the asymptotic symmetry (large gauge symmetry) leads to the memory effects in classical electrodynamics.  
In this section, we argue in the BRST formalism that asymptotic symmetry should be physical one in QED, 
although they are na\"ively parts of the gauge symmetry. 
We first review the covariant quantization of QED in the BRST formalism in subsection \ref{subsec:reviewBRST}. Next, we  look at the asymptotic behaviors of gauge fields in subsection \ref{subsec:as_behavior}. Based on that, we argue that the leading order of the angular components of the gauge fields are physical degrees of freedom in the asymptotic regions, and parts of the large gauge transformations are physical symmetry which transform the physical degrees of freedom nontrivially. In subsection \ref{subsec:leadingsoft}, we review, because it might be useful to read next section,  that the Ward-Takahashi identities for the large transformations actually result in the leading soft photon theorem as \cite{He:2014cra,Campiglia:2015qka}.

\subsection{Review of the covariant quantization in QED}
\label{subsec:reviewBRST}
In this subsection we review the covariant quantization of massive scalar QED in the BRST formalism\footnote{In  QED, the ghost sector is completely decoupled. Nevertheless, we use the BRST formalism to discuss what the physical states are.}, and discuss the symmetries.

The Lagrangian in the Feynman gauge is given by
\ba{\label{Lag:eq}
	\mathcal{L}_{QED}=\mathcal{L}_{EM}+\mathcal{L}_{matter}+\mathcal{L}_{GF}+\mathcal{L}_{FP},
}
where
\ba{
	&\mathcal{L}_{EM}=-\frac{1}{4}F_{\mu\nu}F^{\mu\nu}\ \ \ \  ,\ \ \ \ 
	\mathcal{L}_{matter}=-\frac{1}{2}D_{\mu}\bar{\phi}D_{\mu}\phi-\frac{1}{2}m^2 \bar{\phi}\phi\,,\\
	&\mathcal{L}_{GF}=-\frac{1}{2}\bl\partial^{\mu}A_{\mu}\br^{2}\ ,\
	\mathcal{L}_{FP}=i\partial^{\mu}\bar{c}\partial_{\mu}c.
}
Here, 
$\phi$ is a massive charged scalar field\footnote{In this paper, we consider only a single species of charge for simplicity. The generalization to many species is straightforward.} with charge $e$ where $D_\mu \phi = \partial_\mu \phi - i e A_\mu \phi$, and $c$ is the ghost field.  $\mathcal{L}_{GF}+\mathcal{L}_{FP}$\footnote{We already integrated out the Nakanishi-Lautrup field in $\mathcal{L}_{GF}$.
} is a BRST exact term, which is added so that there is no first class constraint \cite{Kugo:1981hm}. For this Lagrangian, the free Heisenberg equations for gauge fields and the ghost field are $\Box A_{\mu}(x)=0$ and $\Box c(x)=0$, and then the ``general"\footnote{The Fourier expansion automatically gives the following fall-off condition: 
	\ba{
		\lim_{r \to \infty, u: \text{fixed}}c(x)=\mathcal{O}(r^{-1}).\nonumber
	}
	The justification of this fall-off condition is discussed in next subsection.} solutions are given by 
\ba{
	&A_{\mu}(x)=\int \frac{d^3 k}{(2\pi)^3 2\omega_{k}} \bl a_{\mu}(\vec{k}) e^{ikx} +\cre e^{-ikx}\br, \label{A:eq}\\
	&c(x)=\int \frac{d^3 k}{(2\pi)^3 2\omega_{k}} \bl  c(\vec{k}) e^{ikx} +c^{\dagger}(\vec{k}) e^{-ikx}\br, \label{c:eq}
}
where $k^{\mu}$ are massless on-shell momenta. In quantum theory, $a_{\mu}(\vec{k})$ and $c(\vec{k})$ are the annihilation operators for photons and ghost particles with momentum $\vec{k}$, respectively. In particular, the commutation relation of the photon operator is given by 
\begin{align}
[a_{\mu}(\vec{k}), a^\dagger_{\nu}(\vec{k}')] = (2\pi)^3 (2 \omega_k) \eta_{\mu\nu} \delta^{(3)}(\vec{k}-\vec{k}').
\end{align}

The total Lagrangian has a BRST symmetry \cite{Becchi:1975nq, Tyutin:1975qk}. It also has residual symmetries, namely
\ba{\label{rg:eq}
	\delta_{g}\phi(x)=ie\epsilon(x)\phi(x)\ ,\ \delta_{g}\bar{\phi}(x)=-ie\epsilon(x)\bar{\phi}(x)\ ,\ \delta_{g}A_{\mu}(x)=\partial_{\mu}\epsilon(x),
}
with
\ba{
\Box\epsilon(x)=0.
}
These residual symmetries are usually regarded as ``gauge" redundancies, but we will argue in next subsection~\ref{subsec:as_behavior} that parts of them, which are given by large gauge parameters \eqref{LGP}, are physical symmetries, which impose the nontrivial constraints on the $S$-matrices as the Ward-Takahashi identities. Using the Lagrangian \eqref{Lag:eq}, one can find the Noether current associated with \eqref{rg:eq};
\ba{
	j^{\mu}_{\epsilon}(x)=j_{S}^{\mu}(x) +j_{H}^{\mu}(x),
	\label{eq_large_gauge_current}
}
where
\ba{
	&j_{S}^{\mu}(x)=\partial_{\nu}\epsilon(x)F^{\mu\nu}(x)+\partial^{\mu}\epsilon(x)\partial_{\nu}A^{\nu}(x)\,, \\
	&j_{H}^{\mu}(x)=\epsilon(x) j^\mu_{mat}(x)\,, \quad \text{with} \quad 
	j^\mu_{mat}(x)=ie\bigl(D^{\mu}\bar{\phi}(x)\,\phi(x)-\bar{\phi}(x)\,D^{\mu}\phi(x)\bigr).
}
The current is different from \eqref{class_current} by the gauge fixing term. 

We also have a charge associated with the BRST symmetry. It is given by 
\ba{
	Q_{BRST}=\int_{\Sigma} dS^{\mu}\big{[}B(x)\partial_{\mu}c(x)-\partial_{\mu}B(x)c(x) \big{]} , \label{QB:eq}
}
where $\Sigma$ is an arbitrary Cauchy slice and $B(x)=\partial^{\mu}A_{\mu}(x)$. When we take $\Sigma$ as the usual time slice ($t=const.$) and substitute (\ref{A:eq}) and (\ref{c:eq}) into (\ref{QB:eq}), the BRST charge is expressed as  
\ba{
	Q_{BRST}=i\int \frac{d^3 k}{(2\pi)^3 2\omega_{k}} \big{[}c^{\dagger}(\vek)B(\vek)- c(\vec{k})B^{\dagger}(\vek) \big{]},
}
where $B(\vek)$ is defined as 
\begin{align}
B(x) = \int \frac{d^3 k}{(2\pi)^3 2\omega_{k}} \bl  B(\vec{k}) e^{ikx} +B^{\dagger}(\vec{k}) e^{-ikx}\br.
\end{align}
Consequently, if we impose the BRST condition \cite{Kugo:1977zq} on the ghost zero sector of the Hilbert space to extract the physical states, the Gupta-Bleuler condition \cite{Gupta:1949rh,Bleuler:1950cy} is imposed on the physical Hilbert space, 
\ba{\label{GB:eq}
	Q_{BRST}|\psi \rangle=0 \Leftrightarrow B(\vec{k})|\psi \rangle=0,
}
which ensures that the longitudinal modes of gauge fields do not contribute to the dynamics of QED. 
Note that since the ghost field $c(x)$ is decoupled from the gauge fields in the Abelian gauge theory, we can treat $c(x)$ as a Grassmann number function in practice. If gauge fields can be regarded as free fields, $B(\vec{k})$ is related to the annihilation operator of photons as 
$B(\vec{k})= i k^\mu a_\mu (\vec{k})$, and the Gupta-Bleuler condition is written as 
\begin{align}
\label{freeGB:eq}
k^\mu a_\mu (\vec{k}) \ket{\psi}=0\,.
\end{align} 

The subspace annihilated by $Q_{BRST}$ is represented by $\mathcal{H}_{closed}$; 
and it contains the BRST exact subspace $\mathcal{H}_{exact}$ in which states are orthogonal to all states in $\mathcal{H}_{closed}$ since $Q_{BRST}^2=0$. Adding a BRST exact state $\lambda Q_{BRST} \ket{\xi}$, where $\lambda$ is a Grassmann number and $\ket{\xi}$ is any state, to the closed state $\ket{\psi} \in \mathcal{H}_{closed}$ as
\ba{
	|\psi\rangle \rightarrow  \ket{\psi'}=|\psi\rangle+ \lambda Q_{BRST}|\xi\rangle\,,
}
corresponds to a ``gauge transformation", \textit{i.e.} $\ket{\psi}$ and $\ket{\psi'}$ are physically equivalent. 
The true physical space is obtained by identifying such equivalent states and thus given by the cohomology of $Q_{BRST}$:   
\ba{
	\mathcal{H}_{phys}\equiv \mathcal{H}_{closed}/\mathcal{H}_{exact}\,. 
}
Any physical operator $\mathcal{O}$ on $\mathcal{H}_{phys}$ must satisfy the BRST invariant condition, namely 
\ba{\label{phys:eq}
	\delta_{BRST}\mathcal{O}=\big{[}i\lambda Q_{BRST}, \mathcal{O} \big{]}=0\,.
}
If we define the ``physical symmetry" as the symmetry whose charge acts on the physical states nontrivially, then the BRST transformation is not the physical symmetry but a ``gauge symmetry" since the BRST charge acts trivially on physical states by the BRST condition.

\subsection{Asymptotic gauge fields as the physical Cauchy data}\label{subsec:as_behavior}
In this subsection, we consider the asymptotic behaviors of gauge fields to identify the physical degrees of freedom in the asymptotic regions. 
We then argue that the large gauge transformations, which are parts of the symmetry \eqref{rg:eq}, constitute a true physical symmetry, \textit{i.e.}, they are \textit{not} gauge redundancies. 

In order to investigate the asymptotic behaviors near future null infinity $\mathscr{I}^+$, we use the retarded coordinates $(u,r,\Omega^A)$ where $u$ is the retarded time $u=t-r$ and $\Omega^A$ $(A=1,2)$ are (arbitrary) coordinates on the unit two-sphere as in section~\ref{sec:charge conservation}.
In the coordinates, the Minkowski line element is written as
\ba{
	ds^{2}=-du^{2}-2dudr+r^{2}\gamma_{AB}d\Omega^{A}d\Omega^{B}
}
where $\gamma_{AB}(\Omega)$ is the metric on the unit two-sphere. 


At the asymptotic region $\mathscr{I}^+$,
the radiation fields $A_\mu(x)$ would be well approximated by the free field which has the plane wave expansion \eqref{A:eq}. At large $r$ with fixed $u$, one can find the following asymptotic behavior by using the saddle point approximation (see e.g. \cite{Strominger:2017zoo})
\ba{\label{saddle:eq}
	A_{\mu}(x)=-\frac{i}{8\pi^{2}r}\int_{0}^{\infty}d\omega\big{[} a_{\mu}(\omega\hat{x}) e^{-i\omega u}-(h.c.)\big{]}+\mathcal{O}(r^{-2} \log r).
}
 Accordingly, each component of  the gauge field in the $(u,r,\Omega^B)$ coordinates is obtained as follows: 
\ba{
	&A_{u}(x)=\frac{A_{u}^{(1)}(u,\Omega)}{r}+\mathcal{O}(r^{-2} \log r),\nonumber\\
	&A_{r}(x)=\frac{A_{r}^{(1)}(u,\Omega)}{r}+\mathcal{O}(r^{-2} \log r),\label{Asaddle:eq}\\
	&A_{B}(x)=A_{B}^{(0)}(u,\Omega)+\mathcal{O}(r^{-1} \log r)\nonumber
}
with
\ba{
	&A_{u}^{(1)}(u,\Omega)=-\frac{i}{8\pi^{2}}\int_{0}^{\infty}d\omega\big{[} a_{u}(\omega \hat{x}) e^{-i\omega u}-(h.c.)\big{]},\nonumber\\
	&A_{r}^{(1)}(u,\Omega)=-\frac{i}{8\pi^{2}}\int_{0}^{\infty}d\omega\ \big{[} a_{r}(\omega\hat{x}) e^{-i\omega_{k}u}-(h.c.)\big{]},\label{Asaddle_op:eq}\\
	&A_{B}^{(0)}(u,\Omega)=-\frac{i}{8\pi^{2}}\int_{0}^{\infty}d\omega\ \big{[}a_{B}(\omega\hat{x}) e^{-i\omega u}-(h.c.)\big{]},\nonumber
}
where each annihilation operator is defined as 
 \ba{
a_{u}(\vec{k})=a_{t}(\vec{k})\ ,\ a_{r}(\vec{k})=q^{\mu}a_{\mu}(\vec{k})\ ,\  a_{B}(\vec{k})=\frac{\partial \hat{x}^{i}}{\partial \Omega^{B}}a_{i}(\vec{k})\,, \quad q^{\mu}\equiv (1,\hat{x})\,.
 }
Therefore, we can say that the Fourier expansions of $A_{\mu}$ \eqref{A:eq} automatically give the following fall-off conditions: 
\ba{\label{fall:eq}
	\lim_{r \to \infty}A_{r}=\mathcal{O}(r^{-1})\ ,\ \lim_{r \to \infty}A_{u}=\mathcal{O}(r^{-1})\ ,\ \lim_{r \to \infty}A_{B}=\mathcal{O}(1).
}

We now see that the leading $\mathcal{O}(1)$ components $A_B^{(0)}$ constitute the Cauchy data on the future null infinity. In other words, the two components correspond to the two physical degrees of freedom of photons. 
As we have already noted in the previous subsection, we have the BRST symmetry as a ``gauge symmetry". The gauge fields $A_\mu$ transform under the BRST transformations as follows:
\ba{
	&\delta_{BRST}A_{\mu}(x)=\big{[}i\lambda Q_{BRST}, A_{\mu}(x) \big{]}=\lambda\partial_{\mu}c(x)\,.
	\label{eq:gauge_BRST}
}
Since the ghost field satisfies the free equation of motion 
\ba{
	\Box c(x)=0,
	\label{eom:eq}
}
we may use the mode expansion \eqref{c:eq}. However, it is not the general solution of \eqref{eom:eq} in the following sense. 
By using the saddle point approximation as in the calculation of \eqref{saddle:eq}, the ghost field is expanded around $\mathscr{I}^+$ as 
\ba{\label{clarger:eq}
	c(x)=-\frac{i}{8\pi^{2}r}\int_{0}^{\infty}d\omega_{k}\big{[}c(\omega_{k}\hat{x}) e^{-i\omega_{k}u}-(h.c.)\big{]}+\mathcal{O}(r^{-2} \log r).
}
Namely, the expansion (\ref{c:eq}) only describes the solutions with the fall-off condition:
\ba{\label{cfall:eq}
	\lim_{r \to \infty, u: \text{fixed}}c(x)=\mathcal{O}(r^{-1}).
}
Supposing that the ghost field satisfies this fall-off condition,  
we can see that  the leading components $A^{(0)}_{B}(u,\Omega)$ of the gauge fields at the null infinity are invariant under the BRST transformation  \eqref{eq:gauge_BRST}, namely 
\ba{
	\delta_{BRST}A^{(0)}_{B}(u,\Omega)=0.
}
This is because $\partial_B c(x)$ does not have any $\mathcal{O}(1)$ term at the future null infinity due to the fall-off \eqref{cfall:eq}. 
Thus, $A^{(0)}_B$ would be regarded as the physical operators in the sense of \eqref{phys:eq}. They create and annihilate quantum excitations of radiation fields on the null infinity.

In consequence, one can find that  parts of the transformations \eqref{rg:eq} are physical symmetry. In fact, if one takes the form \eqref{LGP} as the parameter function $\epsilon(x)$,  the Cauchy data $A^{(0)}_B$ transform under \eqref{rg:eq} as 
\ba{
 \delta_{g}A_{B}(u, \Omega)=\partial_{B}\epsilon^{(0)}(\Omega). \label{ast}
}
Thus, the large ``gauge" transformations changes the physical operators; in this sense they should be regarded as physical symmetry.   

Here, we remark that  
one can theoretically take the large ``gauge" transformations as gauge redundancies by extending the BRST transformations. 
If one allows the ghost field to take $\mathcal{O}(1)$ values at the null infinity, the transformation \eqref{ast} is regarded as the extended BRST exact variation. One may define the ``physical Hilbert space" as the extended BRST cohomology; in other words, one may require that the ``physical states" should also be singlet under the large gauge transformations. 
However, we know that the large gauge charges generally take nonzero values in the classical scattering events in nature as seen in section~\ref{sec:charge conservation} and appendix~\ref{sec:app_a}.  
Thus, in this extended BRST cohomology, the dynamics of QED is too restricted (or too trivial), \textit{i.e.}, sectors with nonzero charges are ignored. Hence, if we want to have the quantum theory describing scatterings with the electromagnetic memories, we should \textit{not} regard the large gauge transformations as gauge redundancies. This is a justification for the fall-off condition \eqref{cfall:eq} and the conclusion that large ``gauge" transformations \eqref{ast} are physical symmetry.

\subsection{Leading soft photon theorem}\label{subsec:leadingsoft}
We have shown the large gauge transformations act nontrivially on asymptotic states and as a result they are physical symmetries. 
In this subsection we reconfirm (in our notation) by following \cite{Campiglia:2015qka} (see also \cite{Kapec:2015ena}) that the Ward-Takahashi identity for the symmetry is equivalent to the leading soft photon theorem in massive QED (the argument for massless QED is in \cite{He:2014cra}). 
This computation is probably useful to read next section.

The Noether current for the symmetry takes the expression \eqref{eq_large_gauge_current} with parameter $\epsilon(x)$ given by \eqref{LGP}. 
As we saw in section~\ref{sec:charge conservation}, the associated Noether charge is asymptotically conserved, and it consists of soft part $Q_{S}^{\text{lead},\pm}[\epsilon^{(0)}]$ and hard part $Q_{H}^{\text{lead}, \pm}[\epsilon^{(0)}]$. For simplicity, we concentrate on future infinities and omit the analysis for past infinities because it is just a repeat of the similar argument. 

The soft part $Q_{S}^{\text{lead},+}[\epsilon^{(0)}]$ is defined on the future null infinity $\mathscr{I}^+$ and given by
\begin{align}
Q_{S}^{\text{lead}, +}[\epsilon^{(0)}] \equiv \int_{\mathcal{I}^+}dud^{2}\Omega \sqrt{\gamma} \lim_{r\to \infty} \left[r^{2} j_{S}^{r}(u,r,\Omega)\right]\,. 
\end{align}
We then have the expansion 
\begin{align}
r^2   j_{S}^{r}
=\lim_{r\to \infty} \log\frac{2 r}{|u|}\,\partial_{u}A_{r}^{(1)} \Delta_{\mathrm{S}^2} \epsilon^{(0)}
-\gamma^{AB}\partial_{u}A_{B}^{(0)}\partial_{A}\epsilon^{(0)}+\mathcal{O}(r^{-1}\log r)\,, 
\end{align}
because the gauge parameter $\epsilon(x)$ is expanded, as given in eq.~\eqref{epsilon_exp}, as 
\begin{align}
\label{epsilon_exp_ur}
\epsilon(u, r, \Omega) = \epsilon^{(0)}(\Omega)+ \frac{u \log{\frac{2 r}{|u|}}}{2 r} \Delta_{\mathrm{S}^2}\epsilon^{(0)}(\Omega) + \mathcal{O}(r^{-1})\,, 
\end{align}
and gauge fields are expanded as \eqref{Asaddle:eq}. 
The soft charge operator $Q_{S}^{\text{lead}, +}$ is thus given by 
\ba{
	Q^{\text{lead},+}_{S}=Q^{T+}_{S}+Q^{L+}_{S} 
} 
with
\ba{
	Q^{T+}_{S}&=-\int_{\mathcal{I}^+}dud^{2}\Omega\sqrt{\gamma}\gamma^{AB}\partial_{u}A^{(0)}_{B}\partial_{A}\epsilon^{(0)} ,
	\label{LCharge}\\
	Q^{L+}_{S}&=\lim_{r \to \infty}  \int_{\mathcal{I}^+}dud^{2}\Omega\sqrt{\gamma} \log\frac{2 r}{|u|}\partial_{u}A_{r}^{(1)}  \Delta_{\mathrm{S}^2} \epsilon^{(0)}.
}%
The expression of $Q^{T+}_{S}$ agrees with the soft charge \eqref{cl_lead_soft_charge} in the classical case. 
It actually creates or annihilates soft photons \cite{He:2014cra}; Using eq.\eqref{Asaddle_op:eq}, eq.~\eqref{LCharge} can be rewritten as\footnote{The past soft charge $Q^{T-}_{S}$ is defined similarly, and takes the same expression as eq.~\eqref{eq:soft_t_op}. }
\begin{align}
\label{eq:soft_t_op}
Q^{T+}_{S}=\frac{1}{8 \pi}  \lim_{\omega \to 0} \int d^2 \Omega \sqrt{\gamma} \gamma^{AB} \left[\omega a_B(\omega \hat{x}(\Omega)) +\omega a_B^\dagger(\omega \hat{x}(\Omega))\right]\partial_A \epsilon^{(0)}(\Omega)\,.
\end{align}
However, we also have extra divergent term $Q^{L+}_{S}$. The reason why there is no such a term in the classical case is that we impose the Lorenz gauge condition $\partial_\mu A^\mu=0$ which leads to $\partial_u A_{r}^{(1)}=0$. Correspondingly, also at the quantum level, $Q^{L+}_{S}$ vanishes in the physical $S$-matrix elements as follows.  
Using eq.~\eqref{Asaddle_op:eq}, an $S$-matrix element including $\partial_u A_{r}^{(1)}$ is computed as
\begin{align}
\label{eq:long_S}
&\outst \partial_u A_{r}^{(1)} (u, \Omega)\,\mathcal{S} \inst \nn
&=-\frac{1}{8\pi^{2}}\int_{0}^{\infty}d\omega\outst \big{[} \omega q^{\mu}a_{\mu}(\omega \hat{x}) e^{-i\omega u}+\omega q^{\mu} a^{\dagger}_{\mu}(\omega \hat{x}) e^{i\omega u}\big{]}\mathcal{S} \inst,
\end{align}
where $q^\mu = (1,\hat{x})$.
Only longitudinal modes appear in the bracket in the second line. The second term vanishes due to the Gupta-Bleuler condition \eqref{freeGB:eq} and the first term also does by the Ward-Takahashi identity for the global $\mathrm{U}(1)$ symmetry. Thus, eq.~\eqref{eq:long_S} is equal to zero. In other words, the operator $\partial_u A_{r}^{(1)}$ vanishes in the physical amplitude because longitudinal modes of photons do not contribute to the physical scatterings. Thus, we can ignore $Q^{L+}_{S}$. 

Next, we consider hard part $Q_{H}^{\text{lead},+}[\epsilon^{(0)}]$, which is defined on the timelike infinity $i^+$.\footnote{In our notation, the hard charge is defined on the surface $\Sigma_f$ (or $\Sigma_i$) in Fig.~\ref{fig:region} with limits $U\to \infty$ after $T \to \infty$.
Under the limits, the surface $\Sigma_f$ coincide with the constant $\tau$ surface with $\tau \to \infty$. }
It is useful to introduce the coordinates $(\tau, \rho, \Omega^A)$ (see, e.g., \cite{Campiglia:2015qka}). The coordinates are defined as
\ba{
	\tau^{2}=t^{2}-r^{2}\ ,\ \rho=\frac{r}{\sqrt{t^{2}-r^{2}}}.
}
The constant $\tau$-surface is given by three-dimensional hyperbolic space with the curvature radius $\tau$, and the Minkowski line element takes the form
\ba{
	ds^{2}=-d\tau^{2}+\tau^{2}\, h_{\alpha\beta} d \sigma^\alpha d \sigma^\beta, 
}
where $\sigma^\alpha=(\rho, \Omega^A)$ are coordinates of the unit three-dimensional hyperbolic space $\mathbb{H}^3$ with the line element 
\begin{align}
h_{\alpha\beta} d \sigma^\alpha d \sigma^\beta=\frac{d\rho^{2}}{1+\rho^{2}}+\rho^{2}\gamma_{AB}d\Omega^{A}d\Omega^{B} .
\end{align}

If we assume that charged particles can be regarded as free particles near timelike infinity $i^+$,\footnote{We will give comments on this assumption in section~\ref{sec:disc}.} the hard part is given by
\begin{align}
\label{leading_hard_op}
Q_{H}^{\text{lead},+}[\epsilon^{(0)}] = \int_{\mathbb{H}^3} \!\! d^3 \sigma \sqrt{h}\, \epsilon_{\mathbb{H}^3}(\sigma) j_{mat}^{\tau(3)}(\sigma)\,.
\end{align} 
Here, $\epsilon_{\mathbb{H}^3}(\sigma)$ is a limit of large gage parameter $\epsilon(x)$ given by \eqref{LGP} to $i^+$, which is defined as\footnote{In fact, in the coordinates $(\tau,\rho, \sigma)$, Green's function \eqref{green_p} does not depend on  $\tau$. Hence, $\epsilon_{\mathbb{H}^3}(\sigma)=\epsilon(\tau,\sigma)$.} 
\begin{align}
\label{def_e3}
\epsilon_{\mathbb{H}^3}(\sigma)\equiv \lim_{\tau \to \infty} \epsilon(\tau,\rho, \Omega) 
=\int d^2 \Omega' \sqrt{\gamma(\Omega')}\,  G_{\mathbb{H}^3}(\sigma;\Omega') \epsilon^{(0)}(\Omega')
\end{align}
with
\begin{align}
\label{def_G3}
 G_{\mathbb{H}^3}(\sigma;\Omega') = \frac{1}{4\pi \left[-\sqrt{1+\rho^2}+\rho\, \hat{q}(\Omega')\cdot \hat{x}(\Omega)\right]^2}\,.
\end{align}
Besides, $j_{mat}^{\tau(3)}(\sigma)$ in \eqref{leading_hard_op} is the leading coefficient at large $\tau$ of the matter current with the normal ordering, defined as  
\begin{align}
j_{mat}^{\tau(3)}(\sigma) \,\equiv \lim_{\tau \to \infty} \tau^3 :j_{mat}^\tau (\tau, \sigma):\,, 
\end{align}
and it is given by 
\begin{align}
j_{mat}^{\tau(3)}(\sigma) =\frac{em^2}{2(2\pi)^{3}}\left[ b^{\dagger}(\vec{p}) b(\vec{p})-d^{\dagger} (\vec{p}) d(\vec{p}) \right]|_{\vec{p}=m\rho \hat{x}(\Omega)}\,.
\end{align}
See appendix~\eqref{app:asympt_hard} for our convention of the quantization of the scalar field. 
From this expression, one can easily find that hard operator $Q_{H}^{\text{lead},+}[\epsilon^{(0)}]$ acts on a asymptotic future state $\outst$ containing charged particles with momenta $\vec{p}_k =m \rho_k \hat{y}(\tilde{\Omega}_k)$ and charge $e_k$ as\footnote{$\hat{y}(\tilde{\Omega})$ is a unit three-dimensional vector parametrized by spherical coordinates $\tilde{\Omega}^{A}$, and $e_k$ is $+e$ for particles and $-e$ for antiparticles.} 
\begin{align}
\label{eq:lead_hard_+_action}
\outst Q_{H}^{\text{lead},+}[\epsilon^{(0)}]  = \sum_{k\in out} e_k \epsilon_{\mathbb{H}^3}(\rho_k, \tilde{\Omega}_k) \outst\,.
\end{align}
Similarly,  past hard charge $Q_{H}^{\text{lead},-}[\epsilon^{(0)}]$ acts on asymptotic past states as
\begin{align}
\label{eq:lead_hard_-_action}
Q_{H}^{\text{lead},-}[\epsilon^{(0)}] \inst= \sum_{k\in in} e_k \epsilon_{\mathbb{H}^3}(\rho_k, \tilde{\Omega}_k) \inst\,.
\end{align}

As explained in section~\ref{sec:charge conservation}, the leading charges associated with the large gauge transformations are ``asymptotically conserved". Therefore, we should have the following Ward-Takahashi identity for the physical $S$-matrix:  
\ba{\label{acl}
\outst\left[\bl Q^{\text{lead},+}_{S}+Q^{\text{lead},+}_{H}\br \mathcal{S}-\mathcal{S}\bl Q^{\text{lead},-}_{S}+Q^{\text{lead},-}_{H}\br\right]\inst =0.
}
It is known \cite{He:2014cra, Campiglia:2015qka} that 
this identity is equivalent to the leading soft photon theorem \cite{Yennie:1961ad,Weinberg:1965nx}:
\begin{align}
\label{eq:leading_soft_thrm}
\lim_{\omega \to 0} \outst \omega a_B(\omega \hat{x}) \mathcal{S} \inst 
= \left[\sum_{k\in out} \frac{e_k \vec{p}_k \cdot \partial_B \hat{x}}{p_k \cdot q}-\sum_{k\in in} \frac{e_k \vec{p}_k \cdot \partial_B \hat{x}}{p_k \cdot q}\right]
\outst \mathcal{S} \inst\,, 
\end{align}
where $q^\mu=(1,\hat{x})$. 
In fact, integrating  the l.h.s. of the soft theorem \eqref{eq:leading_soft_thrm} w.r.t. the direction $\hat{x}(\Omega)$ of the momentum of the soft photon, we obtain an $S$-matrix element with the insertion of soft charge \eqref{eq:soft_t_op} as follows:
\begin{align}
\label{eq:ppp}
\lim_{\omega \to 0} \frac{1}{4\pi}\int d^2 \Omega \sqrt{\gamma} \gamma^{AB} \partial_A \epsilon^{(0)} \outst \omega a_B(\omega \hat{x}) \mathcal{S} \inst 
= \outst \left( Q^{lead,+}_{S} \mathcal{S}-\mathcal{S} Q^{lead,-}_{S} \right)\inst \,,
\end{align}
where we have used the fact 
\begin{align}
\lim_{\omega \to 0} \outst \omega a_B(\omega \hat{x}) \mathcal{S} \inst  = - \lim_{\omega \to 0} \outst \mathcal{S}\omega a_B^\dagger(\omega \hat{x}) \inst \,.
\end{align}
The soft theorem \eqref{eq:leading_soft_thrm} equates \eqref{eq:ppp} with 
\begin{align}
\frac{1}{4\pi}\int d^2 \Omega \sqrt{\gamma} \gamma^{AB} \partial_A \epsilon^{(0)} 
\sum_{k} \frac{\eta_k e_k \vec{p}_k \cdot \partial_B \hat{x}}{p_k \cdot q}\outst \mathcal{S} \inst\,,
\end{align}
where we have introduced the symbol $\eta_k$ which is $+1$ ($-1$) for $k \in out$ ($k \in in$).   
Performing a partial integration and using the formula
\begin{align}
\nabla_A \left[\gamma^{AB}  \frac{\vec{p}_k \cdot \partial_B \hat{x}(\Omega)}{p_k \cdot q(\Omega)} \right]
=4\pi G_{\mathbb{H}^3}(\rho_k, \tilde{\Omega}_k;\Omega)-1 \quad \text{with} \quad \vec{p}_k\equiv m \rho_k \hat{y}(\tilde{\Omega}_k)\,,
\end{align}
we then have
\begin{align}
&\frac{1}{4\pi}\int d^2 \Omega \sqrt{\gamma} \gamma^{AB} \partial_A \epsilon^{(0)} 
\sum_{k} \frac{\eta_k e_k \vec{p}_k \cdot \partial_B \hat{x}}{p_k \cdot q}= - \sum_{k} \eta_k e_k \int d^2 \Omega \sqrt{\gamma}  \epsilon^{(0)}(\Omega) \left[ G_{\mathbb{H}^3}(\rho_k, \tilde{\Omega}_k;\Omega)-\frac{1}{4\pi}\right]\nn
&=-\sum_k \eta_k e_k \epsilon_{\mathbb{H}^3} (\rho_k, \tilde{\Omega}_k) +\frac{1}{4\pi}\left(\sum_k \eta_k e_k\right)\int d^2 \Omega \sqrt{\gamma}  \epsilon^{(0)}\,.
\end{align}
Since $\sum_{k} \eta_k e_k=0$ due to the total electric charge conservation, we finally obtain 
\begin{align}
\eqref{eq:ppp} = -\sum_k \eta_k e_k \epsilon_{\mathbb{H}^3} (\rho_k, \tilde{\Omega}_k) \outst \mathcal{S} \inst
= -\outst\left(  Q^{\text{lead},+}_{H} \mathcal{S}-\mathcal{S} Q^{\text{lead},-}_{H} \right)\inst 
\end{align}
where we have used \eqref{eq:lead_hard_+_action} and \eqref{eq:lead_hard_-_action}. 
Therefore, we have reconfirmed that we can obtain  the Ward-Takahashi identity \eqref{acl} from the soft theorem \eqref{eq:leading_soft_thrm}, 
and vice versa because  \eqref{acl} holds for any $\epsilon^{(0)}$.

\section{Subleading charges in massive QED}\label{sec:subleading}
In subsection~\ref{subsec:leadingsoft}, we have confirmed that the leading soft theorem is equivalent to the Ward-Takahashi identity for the large gauge transformation. The similar analysis for the subleading soft theorem was done in \cite{Lysov:2014csa, Campiglia:2016hvg, Conde:2016csj} for massless QED. In \cite{Campiglia:2016hvg, Conde:2016csj} it was found that the symmetries are nothing but the large gauge transformations\footnote{The large gauge transformations are slightly different in two papers \cite{Campiglia:2016hvg, Conde:2016csj}. In \cite{Campiglia:2016hvg}, the gauge parameter is $\mathcal{O}(r)$ at $\mathscr{I}^+$, and thus the generator is divergent but includes the subleading finite part, which is relevant to the subleading soft theorem. On the other hand, in \cite{Conde:2016csj}, it  is shown that the subleading part of $\mathcal{O}(1)$ gauge parameter is related to the subleading soft theorem. Our argument is similar to the latter, although the gauge fixing condition is different.}.

We extend the discussions to massive QED, and obtain the expression of the charges associated with the subleading soft theorem. 
First, we review the soft part of the  subleading charges along the work \cite{Lysov:2014csa} in subsection~\ref{sec:subsoft}. In subsection~\ref{sec:sub_hard_part}, we next derive the expression of the hard part of the subleading charges defined on the future (or past) timelike infinity.

\subsection{Subleading soft photon theorem and the soft part of the subleading charges}\label{sec:subsoft}

Like the leading soft theorem \eqref{eq:leading_soft_thrm}, the subleading soft photon theorem gives the following relation between an amplitude containing a soft photon and an amplitude without that: 
\ba{\label{sub soft theorem}
	\lim_{\omega \to 0}\outst (1+\omega\partial_{\omega})a_{B}(\omega \hat{x})\mathcal{S}\inst=S^{(\text{sub})}_B \outst \mathcal{S}\inst 
	\quad \text{with} \quad S^{(\text{sub})}_B \equiv -i \sum_k \frac{e_k q^{\mu} J_{\mu B}^{k}}{p_{k}\cdot q}\,,
}
where the sum in $S^{(\text{sub})}_B$ is taken for all of the incoming and  outgoing charged particles which are labeled by $k$, and $J^{k}_{\mu\nu}$ is the total angular momentum operator of $k$-th particle (with momentum $\vec{p}_k$ and charge $e_k$) defined as 
\begin{align}
J^k_{\mu\nu}=-i \left( p_{k\mu}\frac{\partial}{\partial p_{k}^{\nu}}-p_{k\nu}\frac{\partial}{\partial p_{k}^{\mu}}\right)\,,
\end{align}
and $q^\mu=(1,\hat{x})$ represents the direction of the soft photon.
As seen in the leading theorem \eqref{eq:leading_soft_thrm}, $a_{B}(\omega \hat{x})$ creates $1/\omega$ divergence in the soft limit $\omega$. 
The factor $(1+\omega\partial_{\omega})$ in l.h.s of \eqref{sub soft theorem} removes the leading divergence because $(1+\omega\partial_{\omega})(1/\omega)=0$.

It was argued in \cite{Lysov:2014csa} that the subleading soft theorem \eqref{sub soft theorem} can be interpreted as a Ward-Takahashi identity of asymptotic symmetries, like the leading soft theorem reviewed in section~\ref{subsec:leadingsoft}. 
The subleading soft theorem is a quantum realization of the subleading memory effect \eqref{sub_charge_consv}, and it is written as 
\begin{align}
\label{eq:subWT}
\outst\left[\bl Q^{\text{sub},+}_{S}+Q^{\text{sub},+}_{H}\br \mathcal{S}-\mathcal{S}\bl Q^{\text{sub},-}_{S}+Q^{\text{sub},-}_{H}\br\right]\inst =0.
\end{align}

As noted below eq.~\eqref{eq:sub_soft_cl}, the soft part is given by
\begin{align}
Q^{\text{sub},+}_S = -\frac12 \int_{\mathscr{I}^+}\!\!\!du d^2 \Omega \sqrt{\gamma}  \epsilon^{(0)} u \partial_u \Delta_{\mathrm{S}^2}  \nabla^B A^{(0)}_B,
\end{align}
and using eq.~\eqref{Asaddle_op:eq}, it is further written as 
\begin{align}
Q^{\text{sub},+}_S =-\frac{i}{16\pi} \lim_{\omega \to 0} \int d^2 \Omega \sqrt{\gamma}\, \Delta_{\mathrm{S}^2} \epsilon^{(0)}
\nabla^B \left[  
(1+\omega \partial_\omega)(a_B(\omega \hat{x})-a_B^{\dagger}(\omega \hat{x}))
\right]\,.
\end{align}
The soft part of the past charge, $Q^{\text{sub},-}_S$, also takes the same expression. 
Hence, $Q^{\text{sub},\pm}_S$ contains $(1+\omega \partial_\omega)[a_B(\omega \hat{x})-a_B^{\dagger}(\omega \hat{x})]$ which corresponds to the subleading soft photons. 
The subleading soft theorem \eqref{sub soft theorem} thus states that 
\begin{align}
\label{subsoft}
\outst (Q_S^{\text{sub},+} \mathcal{S} -\mathcal{S} Q_S^{\text{sub},-} )\inst =
-\frac{1}{8\pi} \int d^2 \Omega \sqrt{\gamma}\, \Delta_{\mathrm{S}^2} \epsilon^{(0)} \, \sum_k
\nabla^B \left[\frac{e_k  q^\mu \partial_B \hat{x}^i}{p_k\cdot q} J^k_{\mu i}\right]\outst \mathcal{S}\inst\,.
\end{align}
If operators $Q^{\text{sub},\pm}_H$ exist such that the r.h.s. of \eqref{subsoft} is equal to 
\begin{align}
-\outst (Q_H^{\text{sub},+} \mathcal{S} -\mathcal{S} Q_H^{\text{sub},-} )\inst\,,
\end{align}
we can establish the Ward-Takahashi identity \eqref{eq:subWT}. 

For massless QED, such hard operators $Q^{\text{sub},\pm}_H$ were obtained \cite{Lysov:2014csa,Campiglia:2016hvg,Conde:2016csj}, where the operators are defined on the future and past null infinities $\mathscr{I}^\pm$. 
What we want to do is to obtain the expression of  $Q^{\text{sub},\pm}_H$ for massive charged particles. This is the goal of next subsection.

\subsection{Hard part of the subleading charges}\label{sec:sub_hard_part}

Unlike massless QED, 
$Q^{\text{sub},\pm}_H$ is an operator on timelike infinities $i^\pm$ acting on the asymptotic states of massive particles. Thus, like the leading case \eqref{leading_hard_op}, it should be expressed as an integral over three-dimensional hyperbolic space $\mathbb{H}^3$ with gauge parameter $\epsilon_{\mathbb{H}^3}(\sigma)$ on the space. 
We now obtain such an expression for  the future part $Q^{\text{sub},+}_H$. 

First, let us parametrize an on-shell momentum by $(p,\tilde{\Omega}^{ A})$ as  $p^\mu=(E_p, p \hat{y}(\tilde{\Omega}))$ where $E_p= \sqrt{p^2+m^2}$ and $\hat{y}\cdot \hat{y}=1$. Using this parametrization, the angular momentum operators are expressed as 
\begin{align}
J_{0i}&=i \left[\hat{y}^i E_p \partial_p +\frac{E_p}{p}\tilde{\gamma}^{AB} (\tilde{\partial}_{A} \hat{y}^i) \partial^{\prime}_{B} \right]\,
\\
J_{ij}&=-i\left[\hat{y}^i (\tilde{\partial}_{A} \hat{y}^j) -\hat{y}^j(\tilde{\partial}_{A} \hat{y}^i)\right]\tilde{\gamma}^{AB} \partial^{\prime}_{B},
\end{align}
where $\tilde{\partial}_{A}\equiv\frac{\partial}{\partial \tilde{\Omega}^{ A}}$ is the derivative w.r.t. the direction of on-shell momentum of massive particle. Here, $\tilde{\gamma}^{AB}$ is the inverse of the induced metric $\tilde{\gamma}_{AB} \equiv (\tilde{\partial}_{A} \hat{y}) \cdot (\tilde{\partial}_{B} \hat{y})$.
Note that if we parametrize the on-shell momentum as $\vec{p}= m \rho \vec{y}(\tilde{\Omega})$, $E_p=m\sqrt{1+\rho^2}$, the angular momentum operators also can be represented as
\begin{align}
J_{0i}&=i \sqrt{1+\rho^2} \left[\hat{y}^i \partial_\rho +\frac{1}{\rho}\tilde{\gamma}^{AB}(\tilde{\partial}_{A} \hat{y}^i)\, \tilde{\partial}_{B} \right],\label{eq:J0i}
\\
J_{ij}&=-i\left[\hat{y}^i (\tilde{\partial}_{A} \hat{y}^j) -\hat{y}^j(\tilde{\partial}_{A} \hat{y}^i)\right]\tilde{\gamma}^{AB} \tilde{\partial}_{B}.\label{eq:Jij}
\end{align}

We then define the following operator $Q_B^{\text{sub}}(\Omega)$ with angular index $B$ as 
\begin{align}\label{QB}
Q_B^{\text{sub}}(\Omega) =\frac{e}{2} \int &\frac{d^3p}{(2\pi)^3 2E_p} \frac{q^\mu(\Omega) \partial_B \hat{x}^i(\Omega)}{p\cdot q(\Omega)}\nn
&\times[(J_{\mu i}b^\dagger(\vec{p})) b(\vec{p})- b^\dagger(\vec{p}) (J_{\mu i} b(\vec{p}))-
(J_{\mu i}d^\dagger(\vec{p})) d(\vec{p})+ d^\dagger(\vec{p}) (J_{\mu i} d(\vec{p}))
].
\end{align}
One can confirm, by performing some partial integrations\footnote{These partial integrations involve not only the creation and annihilation operators but also soft factors and the integration measure.}, that the first term and the second term in \eqref{QB} are the same when they act on the physical states. The third term and the forth term are also the same.
Accordingly, one can find that $Q_B^{\text{sub}}$ 
acts on the 1-particle state as
\begin{align}
 Q_B^{\text{sub}}(\Omega) \ket{p_k} &= e_k \frac{q^\mu(\Omega) \partial_B \hat{x}^i(\Omega)}{p_k\cdot q(\Omega)} J^k_{\mu i}\ket{p_k},\\
\bra{p_k} Q_B^{\text{sub}}(\Omega) &= -e_k \frac{q^\mu(\Omega) \partial_B \hat{x}^i(\Omega)}{p_k\cdot q(\Omega)} J^k_{\mu i} \bra{p_k}.
\end{align}
Therefore, if one defines the hard charge operator as 
\begin{align}
\label{eq:sub_hard_1}
Q_H^{\text{sub},\pm}=-\frac{1}{8\pi} \int d^2 \Omega \sqrt{\gamma}\, \Delta_{\mathrm{S}^2} \epsilon^{(0)} \,
\nabla^B Q_B^{sub}(\Omega)\,,
\end{align}
it satisfies the desired property
\begin{align}
\outst (Q_H^{\text{sub},+} \mathcal{S} -\mathcal{S} Q_H^{\text{sub},-} )\inst=
\frac{1}{8\pi} \int d^2 \Omega \sqrt{\gamma}\, \Delta_{\mathrm{S}^2} \epsilon^{(0)} \, \sum_k
\nabla^B \left[\frac{e_k  q^\mu \partial_B \hat{x}^i}{p_k\cdot q} J^k_{\mu i}\right]\outst \mathcal{S}\inst\,.
\end{align}

Next, we now express $Q_B^{\text{sub}}$ in terms of the local matter current of charged particles in the asymptotic region $i^+$. 
The matter current $j^{mat}_\mu$ asymptotically decays as $\mathcal{O}(\tau^{-3})$ with $\tau$-dependent oscillations. 
Assuming that the charged scalar is free in the asymptotic region, one can extract $\tau$-independent finite parts of $j^{mat}_\mu$ (see appendix~\ref{app:asympt_hard}) 
as 
\begin{align}
I_{\alpha}^{mat}(\tilde{\sigma})&\equiv \lim_{\tau\to \infty}\left(\frac{1}{4m^2}\partial_{\tau}^{2}+1\right) \tau^3 :j_{\alpha}^{mat}(\tau,\tilde{\sigma}):\\
&=\frac{iem}{4(2\pi)^{3}} \left[\partial_{\alpha}b^{\dagger} (\vec{p})b(\vec{p})-b^{\dagger}(\vec{p})\partial_{\alpha} b(\vec{p})-\partial_{\alpha}d^{\dagger}(\vec{p})d(\vec{p})+d^{\dagger}(\vec{p})\partial_{\alpha} d(\vec{p})
\right]|_{\vec{p}=m \rho \hat{y}(\tilde{\Omega})}\,,
\end{align}
where $\tilde{\sigma}^\alpha =(\rho, \tilde{\Omega}^A)$ are the coordinates on $\mathbb{H}^3$. 
In addition, using this and also eqs.~\eqref{eq:J0i}, \eqref{eq:Jij}, one can obtain the following equations: 
\begin{align}
&[(J_{0 i}b^\dagger(\vec{p})) b(\vec{p})- b^\dagger(\vec{p}) (J_{0 i} b(\vec{p}))-
(J_{0 i}d^\dagger(\vec{p})) d(\vec{p})+ d^\dagger(\vec{p}) (J_{0 i} d(\vec{p}))]|_{\vec{p}=m\rho \hat{y}(\tilde{\Omega})}\nn
&=\frac{4(2\pi)^{3}}{e m}\sqrt{1+\rho^2}[\hat{y}^{i}I_{\rho}^{mat}(\rho, \tilde{\Omega})+\frac{1}{\rho}\tilde{\gamma}^{AB}\tilde{\partial}_{A} \hat{y}^{i} I_{B}^{mat}(\rho, \tilde{\Omega})]\label{0i}\ ,
\end{align}
\vspace{-1cm}
\begin{align}
&[(J_{ij}b^\dagger(\vec{p})) b(\vec{p})- b^\dagger(\vec{p}) (J_{ij} b(\vec{p}))-
(J_{ij}d^\dagger(\vec{p})) d(\vec{p})+ d^\dagger(\vec{p}) (J_{ij} d(\vec{p}))]|_{\vec{p}=m\rho \hat{y}(\tilde{\Omega})}\nonumber\\
&\ =-\frac{4(2\pi)^{3}}{e m}
(\hat{y}^{i}\tilde{\partial}_{A} \hat{y}^{j}-\hat{y}^{j}\tilde{\partial}_{A} \hat{y}^{i}) \tilde{\gamma}^{AB} I_B^{mat}(\rho, \tilde{\Omega}).\label{ij}
\end{align}
From these equations, \eqref{QB} can be rewritten as
\begin{align}
&Q_B^{\text{sub}}(\Omega) = \int_{\mathbb{H}^3}\!\! d^3 \tilde{\sigma} \sqrt{\tilde{h}} \left[
\frac{\sqrt{1+\rho^2} \partial_B \hat{x}(\Omega) \cdot \hat{y}(\tilde{\Omega})}{q\cdot Y}I_{\rho}^{mat}(\rho, \tilde{\Omega})\right.\nn
&\left. +\frac{1}{q\cdot Y}
\left\{\frac{\sqrt{1+\rho^2}}{\rho}\partial_B \hat{x} \cdot \tilde{\partial}_{A} \hat{y}
-(\hat{x}\cdot\hat{y})(\partial_B \hat{x} \cdot \tilde{\partial}_{A} \hat{y})
+(\hat{x} \cdot \tilde{\partial}_{A} \hat{y}) (\partial_B \hat{x} \cdot \hat{y})
\right\}\tilde{\gamma}^{ AC} I_{C}^{mat}(\rho, \tilde{\Omega})
\right]\,,
\label{eq:QB_current}
\end{align}
where $d^3 \tilde{\sigma} \sqrt{\tilde{h}} = d\rho d^2 \tilde{\Omega} \frac{\rho^2}{\sqrt{1+\rho^2}}\sqrt{\tilde{\gamma}}$ and $Y^\mu = (\sqrt{1+\rho^2},\rho \hat{y}(\tilde{\Omega}))$.

Therefore,
the hard charge $Q_H^{\text{sub},+}$ can be expressed in terms of the asymptotic matter current $I_{\alpha}^{mat}$ by inserting \eqref{eq:QB_current} into \eqref{eq:sub_hard_1}. 
However, it seems to be unnatural because $Q_H^{\text{sub},+}$ is given by an integral over $S^2$ with parameter function $\epsilon^{(0)}$, not $\epsilon_{\mathbb{H}^3}$. 
Since $Q_H^{\text{sub},+}$ is associated with the large gauge transformation acting on massive particles, it should be written as an integral over the surface at timelike infinity $\mathbb{H}^{3}$ with parameter function $\epsilon_{\mathbb{H}^3}$ on that surface, like the leading case \eqref{leading_hard_op}. 
In fact, after some computations (see appendix~\ref{cal}), one can express $Q_H^{\text{sub},+}$ in such an integral as follows:
\begin{align}
\label{eq:sub_hard_h3}
Q_H^{\text{sub},+}= \frac12\int_{\mathbb{H}^3}\!\! d^3 \sigma \sqrt{h} \frac{\sqrt{1+\rho^2}}{\rho} \left[
\rho^2 h^{\alpha\beta}(\nabla^{(h)}_\alpha \nabla^{(h)}_\rho  \epsilon_{\mathbb{H}^3})I_{\beta}^{mat}
+2 \rho h^{\alpha\beta}(\nabla^{(h)}_\alpha \epsilon_{\mathbb{H}^3})I_{\beta}^{mat}
\right]
,
\end{align}
where $\nabla^{(h)}_\alpha$ denotes the covariant derivative compatible with the metric $h_{\alpha\beta}$ on $\mathbb{H}^3$.

\section{Discussions and Outlook}\label{sec:disc}

In section~\ref{sec:charge conservation}, we have shown that the leading memory effect is nothing but the charge conservation of the large gauge transformations at the leading order in classical electromagnetism. In addition, looking at the subleading order, we can obtain the subleading memory effect. One can extend this analysis to gravity. There is the sub-subleading graviton theorem in perturbative gravity \cite{Cachazo:2014fwa}. Thus, it is expected that, unlike electromagnetism, the contributions from the spacelike infinity vanish even at the sub-subleading order, and we have the sub-subleading memory effect.\footnote{It was shown that the classical gravitational wave produced by a kick of particles has a term corresponding to the sub-subleading memory \cite{Hamada:2018cjj}.}
It is more interesting to work in the (dynamical) blackhole backgrounds \cite{Hawking:2016msc, Chu:2018tzu}. 

The conserved charges in section~\ref{sec:charge conservation} resemble \textit{multipole charges} in \cite{Seraj:2016jxi} (see also its gravitational extension \cite{Compere:2017wrj}). 
The multipole charges are also defined as the Noether charge for the residual large gauge symmetry in the Lorenz gauge. 
In \cite{Seraj:2016jxi}, it was shown that electric multipole moments of charged matters are conserved if we also take account of the contributions from radiations. 
However, the $l$-th multipole charges are associated with the time-independent large gauge parameters $\epsilon(x)$ which behave as $r^{l} Y_{lm}(\Omega)$ at large $r$. 
Thus, they are different from the large gauge parameters that we considered, and we are not sure whether they are related.

In subsection~\ref{subsec:leadingsoft}, we have reviewed that the Ward-Takahashi identity for the large gauge symmetry is equivalent to the leading soft photon theorem. 
In the analysis, we have assumed that we can regard massive particles as free particles in the asymptotic region.  
However, as we saw in the classical case in subsection~\ref{subsec:leadingsoft}, the hard charge $Q^{\text{lead},+}_H$ always contains the contributions from electromagnetic potential created by massive charges. 
This problems has a long history, and related to the infrared divergences in QED \cite{Chung:1965zza, Kulish:1970ut}. 
In \cite{Chung:1965zza, Kulish:1970ut}, asymptotic states are defined by solving the asymptotic dynamics, and it is shown that such asymptotic states do not cause IR divergences in the $S$-matrix at least in simple examples (see also \cite{Ware:2013zja} where similar arguments were done for gravity). 
Recently, the connection between the IR-finite asymptotic states and asymptotic symmetry has been discussed \cite{Mirbabayi:2016axw, Gabai:2016kuf, Kapec:2017tkm, Choi:2017bna, Choi:2017ylo, Carney:2018ygh}. 
As we argued in section~\ref{sec:brst}, large gauge symmetry is actually physical symmetry. 
Thus, we should classify states by the representation of the symmetry.  
It is interesting to consider this subject in the BRST formalism, and we leave this problem for our future work.

In section~\ref{sec:subleading}, we also used the same assumption that particles are free in the asymptotic regions. 
Probably, the use of appropriate asymptotic states is more relevant in the analysis at the subleading order. 
In the derivation of the expression of the ``hard parts", we have assumed the subleading soft theorem and have written down the operator associated with the subleading soft factors. We did not look at the consistency with the expression of the ``hard parts" of the subleading charges derived in the classical case in section~\ref{sec:charge conservation}. 
The ``hard parts" consist of $Q_+^{\text{log}\prime}+Q_f^\text{log}$, and $Q_+^{\text{log}\prime}$ is defined on $\mathscr{I}^+$, while $Q_f^\text{log}$ on $\mathbb{H}^3$ at future timelike infinity. Each of $Q_+^{\text{log}\prime}$ and $Q_f^\text{log}$ is a divergent quantity in the large $U$ limit, although the sum is finite. Thus, we cannot consider them separately, and we think that the finiteness of $Q_+^{\text{log}\prime}+Q_f^\text{log}$ is ensured only when we take account of the asymptotic interactions in QED. We hope to come back to this issue in the non-asymptotic future.

\section*{Acknowledgement}
We thank Satoshi Yamaguchi for useful discussions. SS would like to also thank the audience of his seminar at String Theory Group in National Taiwan University, especially Yu-tin Huang, for stimulating discussions. 
The work of SS is supported in part by the Grant-in-Aid for Japan Society for the Promotion of Science (JSPS) Fellows No.16J01004.
\appendix
\section{A concrete example}\label{sec:app_a}
\subsection{Electromagnetic fields of uniformly moving charges}
\label{app_a}
If there is a charge $e$ moving with a constant velocity $\vec{v}$ as 
\begin{align}
\vec{x}=\vec{x}_0+\vec{v}(t-t_0)\,, 
\end{align}
the gauge potential produced by the charge in the Lorenz gauge is 
\begin{align}
A^0(x)=\frac{e}{4\pi\, \ell(x)}\,,\quad
\vec{A}(x)=\frac{e \vec{v}}{4\pi\, \ell(x)}\,,
\end{align}
where
\begin{align}
\label{ell(x)}
\ell(x)=\sqrt{(1-|\vec{v}|^2)(|\vec{x}-\vec{x}_0|^2-[\hat{v}\cdot(\vec{x}-\vec{x}_0)]^2)+(\hat{v}\cdot(\vec{x}-\vec{x}_0)-|\vec{v}| (t-t_0))^2}
\end{align}
with
\begin{align}
\hat{v}=\frac{\vec{v}}{|\vec{v}|}.
\end{align}
At the point $t=T-U, r=T+U$ with large $T$, the electric flux $F_{tr}$ is expanded as 
\begin{align}
F_{tr}(t,r,\Omega)|_{t= T-U, r=T+U}=&-\frac{e(1-|\vec{v}|^2)}{4\pi (1-\vec{v}\cdot\hat{x}(\Omega))^2 T^2}
+\frac{e(1-|\vec{v}|^2)f(U,\Omega;\vec{v},t_0,\vec{x}_0)}{4\pi (1-\vec{v}\cdot\hat{x}(\Omega))^4 T^3}+\mathcal{O}(T^{-4})
\end{align}
with
\begin{align}
f(U,\Omega;\vec{v},t_0,\vec{x}_0)\equiv &2 U (1-|\vec{v}|^2-2|\vec{v}_\bot|^2)
+[1-\vec{v}\cdot\hat{x}(\Omega)-3(1-|\vec{v}|^2)]\vec{x}_0\cdot\hat{x}(\Omega) \nn
&+3[1-\vec{v}\cdot\hat{x}(\Omega)] \vec{v}\cdot \vec{x}_0
+[2(1-|\vec{v}|^2)-2(1-\vec{v}\cdot\hat{x}(\Omega))-|\vec{v}_\bot|^2]t_0\,,
\end{align}
where $\vec{v}_\bot\equiv \vec{v}-[\vec{v}\cdot\hat{x}(\Omega)]\hat{x}(\Omega)$. 
At the point $t=-T+U, r=T+U$ with large $T$, the electric flux $F_{tr}$ is 
\begin{align}
F_{tr}(t,r,\Omega)|_{t= -T+U, r=T+U}=&-\frac{e(1-|\vec{v}|^2)}{4\pi (1+\vec{v}\cdot\hat{x}(\Omega))^2 T^2}
+\frac{e(1-|\vec{v}|^2)f(U,\Omega;-\vec{v},-t_0,\vec{x}_0)}{4\pi (1+\vec{v}\cdot\hat{x}(\Omega))^4 T^3}+\mathcal{O}(T^{-4}).
\end{align}
Thus, the antipodal matching condition eq.~\eqref{em_match} at the leading order  
\begin{align}
F^{+(2)}_{tr}(\Omega)=F^{-(2)}_{tr}(\bar\Omega)
\end{align}
holds where $\bar{\Omega}$ denotes the antipodal point of $\Omega$. 

On the other hand, for eq.~\eqref{green_p}, in the limit that $r\to \infty$ with $u=t-r$ fixed, we have  
\begin{align}
\label{green_exp_sp}
\int d^2 \Omega' \sqrt{\gamma(\Omega')} G(u,r,\Omega;\Omega') Y_{\ell m} (\Omega')
=Y_{\ell m} (\Omega) + \frac{\ell(\ell+1)u \log \frac{|u|}{2r} + s_\ell u}{2r}Y_{\ell m} (\Omega) +\mathcal{O}(r^{-1-\varepsilon})\,,
\end{align}
where $Y_{\ell m} (\Omega)$ are the spherical harmonics, and the coefficients $s_\ell$ are\footnote{$c_0=0, c_1=2$.} 
\begin{align}
s_\ell =\frac{1}{2^\ell} \sum_{j=0}^{\lfloor \ell/2\rfloor} \frac{(-1)^j(2\ell-2j)!}{j!(\ell-j)!(\ell-2j)!}c_{\ell-2j} \quad
\text{with}\quad 
c_n=-1+(-1)^n
+n\left(4\sum_{k=1}^n\frac1k
-2\sum_{k=1}^{\lfloor n/2 \rfloor} \frac1k \right).
\end{align}
As a result, the large gauge parameter $\epsilon(u, r, \Omega)$ has following large-$r$ expansion,
\ba{\label{large-r_ep}
\epsilon(u, r, \Omega) = \epsilon^{(0)}(\Omega)+ \frac{u \log{\frac{2 r}{|u|}}}{2 r} \Delta_{\mathrm{S}^2}\epsilon^{(0)}(\Omega) + \mathcal{O}(r^{-1}).
}
Similarly, in the limit that $r\to \infty$ with $v=t+r$ fixed, it is expanded as 
\begin{align}
\epsilon(v, r, \Omega) = \epsilon^{(0)}(\bar{\Omega})- \frac{v \log{\frac{2 r}{|v|}}}{2 r} \Delta_{\mathrm{S}^2}\epsilon^{(0)}(\bar{\Omega})+ \mathcal{O}(r^{-1}).
\end{align}
Therefore, if we define the coefficients of large-$r$ expansion of $\epsilon(x)$ as
\ba{
&\lim_{r \to \infty, u: \text{fixed}}\epsilon(x)=\epsilon^{(0)}(\Omega)+\epsilon^{(log,+)}(u, \Omega)\frac{\log{r}}{r}+\mathcal{O}(r^{-1}),\\
&\lim_{r \to \infty, v: \text{fixed}}\epsilon(x)=\epsilon^{(0)}(\bar{\Omega})+\epsilon^{(log,-)}(v, \Omega)\frac{\log{r}}{r}+\mathcal{O}(r^{-1}),
}
then $\epsilon^{(log,+)}(u=-2U, \Omega)=\epsilon^{(log,-)}(v=2U, \bar{\Omega})$ holds.
Thus, if we have initially charges $e_n$ moving as 
\begin{align}
\vec{x}^{(n)}=\vec{x}^{(n)}_0+\vec{v}_n(t-t^{(n)}_0)\,, 
\end{align}
$Q_0$ given by eq.~\eqref{Q0_sum} is computed as 
\begin{align}
\label{Q_0_app}
Q_0 = \sum_n & \frac{e_n(1-|\vec{v}_n|^2)}{2\pi T}
\int d^2 \Omega \sqrt{\gamma(\Omega)}\frac{\epsilon^{(0)}(\Omega)}{(1-\vec{v}_n\cdot\hat{x}(\Omega))^4} \left\{
[1-\vec{v}_n\cdot\hat{x}(\Omega)-3(1-|\vec{v}_n|^2)]\vec{x}^{(n)}_0\cdot\hat{x}(\Omega)\right. \nn
&\left. +3[1-\vec{v}_n\cdot\hat{x}(\Omega)] \vec{v}_n\cdot \vec{x}^{(n)}_0
+[2(1-|\vec{v}_n|^2)-2(1-\vec{v}_n\cdot\hat{x}(\Omega))-|\vec{v}_{n\bot}|^2]t^{(n)}_0
\right\}+\mathcal{O}(T^{-2})\,.
\end{align}

\subsection{Computation of memories}
\label{app_comp_memo}
Here, we check the memory effect formulae \eqref{lead_memory} and \eqref{sublead_memory} for a concrete example. 
We consider the following trajectory of a charged particle with charge $e$ such as it first rests at $\vec{x}_0$ and moves with a constant velocity $\vec{v}$ after a time $t_0$: 
\begin{align}
\vec{x}=\vec{x}_0+\Theta(t-t_0)\vec{v}(t-t_0).
\label{trajectory}
\end{align}
We represent the matter current for this trajectory by $j_{mat}^\mu$, which is the source in Maxwell's equation $\partial_\nu F^{\nu\mu}=-j_{mat}^\mu$. 
The retarded electromagnetic field created by this particle is written in the Lorenz gauge $\partial_\mu A^\mu=0$ as 
\begin{align}
\label{a_gauge_field}
&A^0(x)=\Theta(|\vec{x}-\vec{x}_0|-t+t_0)\frac{e}{4\pi |\vec{x}-\vec{x}_0|}
+\Theta(-|\vec{x}-\vec{x}_0|+t-t_0)\frac{e}{4\pi\, \ell(x)}
\,,\\
&\vec{A}(x)=\Theta(-|\vec{x}-\vec{x}_0|+t-t_0)\frac{e \vec{v}}{4\pi\, \ell(x)}\,,
\end{align}
where $\ell(x)$ is given by eq.~\eqref{ell(x)}.

We first consider the charge $Q_f$. It is given by 
\begin{align}
Q_f = \int d^2 \Omega \sqrt{\gamma} \left(r^2 F^{tr} \epsilon \right) |_{t=T+U, r=T-U}.
\end{align}  
At $t=T+U, r=T-U$ with large $T$, electric field $F^{tr}$ is expanded as 
\begin{align}
F^{tr}|_{t=T+U, r=T-U}= \frac{e(1-|\vec{v}|^2)}{4\pi (1-\vec{v} \cdot \hat{x})^2 T^2} + \mathcal{O}(T^{-3})\,.
\end{align}
Since our gauge parameter has the expansion as eq.~\eqref{epsilon_exp}, $Q_f$ is expanded as 
\begin{align}
Q_f= \frac{e(1-|\vec{v}|^2)}{4 \pi}\int d^2 \Omega \sqrt{\gamma} \frac{\epsilon^{(0)}}{(1-\vec{v} \cdot \hat{x})^2}
+\frac{U\log T}{T}\frac{e(1-|\vec{v}|^2)}{4 \pi}\int d^2 \Omega \sqrt{\gamma} \frac{\Delta_{\mathrm{S}^2}\epsilon^{(0)}}{(1-\vec{v} \cdot \hat{x})^2}+\mathcal{O}(T^{-1})\,.
\end{align}
Thus, we have 
\begin{align}
\lim_{T\to \infty}Q_f[\epsilon^{(0)}]&=\frac{e(1-|\vec{v}|^2)}{4 \pi}\int d^2 \Omega \sqrt{\gamma} \frac{\epsilon^{(0)}}{(1-\vec{v} \cdot \hat{x})^2}\,,\\
Q_{f}^\text{log}[\epsilon^{(0)}] &=-\frac{Ue(1-|\vec{v}|^2)}{4 \pi}\int d^2 \Omega \sqrt{\gamma} \frac{\Delta_{\mathrm{S}^2}\epsilon^{(0)}}{(1-\vec{v} \cdot \hat{x})^2}= -U \lim_{T\to \infty}Q_f[\Delta_{\mathrm{S}^2}\epsilon^{(0)}]\,.
\end{align}
Note that $Q_{f}^\text{log}$ diverges in the limit $U\to \infty$ although $Q_{f}^\text{log}+Q_{+}^{\text{log}\prime}$ is finite as we will see later.
Similarly, the charge $Q_i$ is computed as 
\begin{align}
\lim_{T\to \infty}Q_i[\epsilon^{(0)}]&=\frac{e}{4 \pi}\int d^2 \Omega \sqrt{\gamma} \epsilon^{(0)}\,,\\
Q_{i}^\text{log}[\epsilon^{(0)}] &=-\frac{Ue}{4 \pi}\int d^2 \Omega \sqrt{\gamma} \Delta_{\mathrm{S}^2}\epsilon^{(0)}=0\,.
\end{align}

Next, we compute the future null infinity charge $\lim_{T\to \infty}Q_+$ given by \eqref{finite_soft}. 
Since the angular components of the gauge field is expanded as 
\begin{align}
A_B (x)= \Theta(u+\vec{x}_0\cdot\hat{x}-t_0) \frac{e \vec{v}\cdot\partial_B \hat{x}}{4\pi (1-\vec{v}\cdot \hat{x})} + \mathcal{O}(T^{-1})\,,
\end{align}
the charge is given by
\begin{align}
\lim_{T\to \infty}Q_+= \int d^2 \Omega \sqrt{\gamma} \epsilon^{(0)}\gamma^{AB} \nabla_A  \left[\frac{e \vec{v}\cdot\partial_B \hat{x}}{4\pi (1-\vec{v}\cdot \hat{x})}\right]\,.
\end{align}
Noting that the formula 
\begin{align}
\gamma^{AB} \nabla_A  \left[\frac{\vec{v}\cdot\partial_B \hat{x}}{1-\vec{v}\cdot \hat{x}}\right]=
\frac{-2 \vec{v}\cdot\hat{x}}{1-\vec{v}\cdot \hat{x}}+\frac{|\vec{v}|^2-(\vec{v}\cdot\hat{x})^2}{(1-\vec{v}\cdot \hat{x})^2}=1-\frac{1-|\vec{v}|^2}{(1-\vec{v}\cdot \hat{x})^2}\,,
\end{align}
the charge has the form 
\begin{align}
\lim_{T\to \infty}Q_+=\frac{e}{4\pi}\int d^2 \Omega \sqrt{\gamma} \epsilon^{(0)}-\frac{e(1-|\vec{v}|^2)}{4 \pi}\int d^2 \Omega \sqrt{\gamma} \frac{\epsilon^{(0)}}{(1-\vec{v} \cdot \hat{x})^2}
=-\lim_{T\to \infty} (Q_f-Q_i)\,.
\end{align}
This certainly agrees with the leading memory effect \eqref{lead_memory}.

Finally, we compute the subleading charges $Q_+^\text{log}$ and  $Q_+^{\text{log} \prime}$. 
Since we now have 
\begin{align}
\partial_u A^{(0)}_B= \delta(u+\vec{x}_0\cdot\hat{x}-t_0) \frac{e \vec{v}\cdot\partial_B \hat{x}}{4\pi (1-\vec{v}\cdot \hat{x})}\,,
\end{align}
the charge $Q_+^\text{log}$ given by eq.~\eqref{q+log} is computed as 
\begin{align}
Q_+^\text{log}&=-\frac12 \int d^2 \Omega \sqrt{\gamma}  \gamma^{AB}(\vec{x}_0\cdot\hat{x}-t_0) \frac{e \vec{v}\cdot\partial_B \hat{x}}{4\pi (1-\vec{v}\cdot \hat{x})} \nabla_A \Delta_{\mathrm{S}^2} \epsilon^{(0)}\nn
&=\frac{e}{8\pi}\int d^2 \Omega \sqrt{\gamma}\left[
(\vec{x}_0\cdot\hat{x}-t_0)\left(1-\frac{1-|\vec{v}|^2}{(1-\vec{v}\cdot \hat{x})^2}\right)
+\frac{\vec{x}_0\cdot\vec{v}-(\vec{v}\cdot \hat{x})(\vec{x}_0\cdot \hat{x})}{1-\vec{v}\cdot \hat{x}}
\right]\Delta_{\mathrm{S}^2} \epsilon^{(0)}\,.
\end{align}
Note that this does not depend on $U$. 
As shown in \cite{Hamada:2018cjj}, this charge is related to the soft factor in the subleading soft photon theorem. 
The momentum of the charged particle is initially $p^\mu = m (1, 0)$ and finally $p^{\prime \mu} = \omega (1,\vec{v})$ with $\omega = m/\sqrt{1-|\vec{v}|^2}$. 
The angular momentum is initially $J^{\mu\nu}= x_0^\mu p^\nu - x_0^\nu p^\mu$ and finally $J^{\prime \mu\nu}= x_0^\mu p^{\prime\nu} - x_0^\nu p^{\prime\mu}$. They read $p^{u}=m=-q\cdot p$, $p^{\prime u}=\omega (1-\vec{v}\cdot \hat{x})=-q\cdot p'$, $p'_B =\omega \vec{v} \cdot \partial_B \vec{x}$, $J^{u}_{\phantom{u}B}=-p^u\vec{x}_0\cdot \partial_B \vec{x}$ and 
$J^{\prime u}_{\phantom{u}B}=-(\vec{x}_0\cdot\hat{x}-t_0)p'_B-p^{\prime u} \vec{x}_0\cdot \partial_B \vec{x}$ where $q^\mu= (1,\hat{x})$. 
Using  them, we have 
\begin{align}
(\vec{x}_0\cdot\hat{x}-t_0) \frac{\vec{v}\cdot\partial_B \hat{x}}{1-\vec{v}\cdot \hat{x}}=\frac{1}{r} \left[
\frac{J^{\prime u}_{\phantom{u}B}}{q\cdot p'} -\frac{J^u_{\phantom{u}B}}{q\cdot p} \right]. 
\end{align}
Thus, $Q_+^\text{log}$ can also be written as 
\begin{align}
Q_+^\text{log}&=-\frac{e}{8\pi} \int d^2 \Omega \sqrt{\gamma} \lim_{r\to \infty} \left( \frac{rJ^{\prime u A}}{q\cdot p'} -\frac{rJ^{uA}}{q\cdot p}\right)
\nabla_A \Delta_{\mathrm{S}^2} \epsilon^{(0)}\,.
\end{align}

The charge $Q_+^{\text{log}\prime}$ given by \eqref{q+logpr} is computed as follows. 
The radial component $A_r$ in $(u,r, \Omega)$ coordinates is 
\begin{align}
A_r= -\frac{e}{4\pi r}　+\mathcal{O}(r^{-2}),
\end{align}
and we thus have $A^{(1)}_r=-e/(4\pi)$, which does not contribute to the charge $Q_+^{\text{log}\prime}$ because 
\begin{align}
\int  d^2 \Omega \sqrt{\gamma}
A_r^{(1)} \Delta_{\mathrm{S}^2} \epsilon^{(0)} = -\frac{e}{4\pi}\int  d^2 \Omega \sqrt{\gamma}\Delta_{\mathrm{S}^2} \epsilon^{(0)}=0\,.
\end{align}
We also have $C_u^{(1)}=0$, $C_A^{(1)}=0$ and 
\begin{align}
\nabla^B A^{(0)}_B &= \frac{e}{4\pi} \Theta(u+\vec{x}_0\cdot\hat{x}-t_0) \left[1-\frac{1-|\vec{v}|^2}{(1-\vec{v}\cdot \hat{x})^2}\right]\nn
&\quad +\frac{e}{4\pi} \delta(u+\vec{x}_0\cdot\hat{x}-t_0)\left[
\frac{\vec{x}_0\cdot\vec{v}-(\vec{v}\cdot \hat{x})(\vec{x}_0\cdot \hat{x})}{1-\vec{v}\cdot \hat{x}}
\right]\,.
\end{align}
Therefore, 
\begin{align}
Q_+^{\text{log}\prime} &= - \frac{e}{8\pi}\int d^2 \Omega \sqrt{\gamma} (2U+\vec{x}_0\cdot\hat{x}-t_0)\left[1-\frac{1-|\vec{v}|^2}{(1-\vec{v}\cdot \hat{x})^2}\right]\Delta_{\mathrm{S}^2} \epsilon^{(0)}\nn
&\quad-\frac{e}{8\pi}\int d^2 \Omega \sqrt{\gamma}\left[
\frac{\vec{x}_0\cdot\vec{v}-(\vec{v}\cdot \hat{x})(\vec{x}_0\cdot \hat{x})}{1-\vec{v}\cdot \hat{x}}
\right]
\Delta_{\mathrm{S}^2} \epsilon^{(0)}
\,,
\end{align}
and for any $\epsilon^{(0)}(\Omega)$ we have 
\begin{align}
Q_+^{\text{log}\prime} +Q_{f}^\text{log}
&=- \frac{e}{8\pi}\int d^2 \Omega \sqrt{\gamma} (\vec{x}_0\cdot\hat{x}-t_0)\left[1-\frac{1-|\vec{v}|^2}{(1-\vec{v}\cdot \hat{x})^2}\right]\Delta_{\mathrm{S}^2} \epsilon^{(0)}\nn
&\quad-\frac{e}{8\pi}\int d^2 \Omega \sqrt{\gamma}\left[
\frac{\vec{x}_0\cdot\vec{v}-(\vec{v}\cdot \hat{x})(\vec{x}_0\cdot \hat{x})}{1-\vec{v}\cdot \hat{x}}
\right]
\Delta_{\mathrm{S}^2} \epsilon^{(0)}\\
&=-Q_+^\text{log}\,.
\end{align}
This is the subleading memory effect.  

\section{Asymptotic expansion of radiation fields}\label{app:asympt_rad}
We here investigate the large-$r$ expansion of radiation fields in the Lorenz gauge $\partial_\mu A^\mu=0$. 
We suppose that gauge fields are generally expanded as follows:\footnote{The expansion is more general than that in \cite{Strominger:2017zoo}, because we allow $\log r$ terms like the gauge parameter $\epsilon(x)$ [see \eqref{epsilon_exp}]. } 
\begin{align}
A_u&=\frac{\log\frac{|u|}{2r}}{r}C_u^{(1)}(u,\Omega)+\frac{1}{r} A_u^{(1)}(u,\Omega)+\frac{\log\frac{|u|}{2r}}{r^2}C_u^{(2)}(u,\Omega)+\frac{1}{r^2}A_u^{(2)}(u,\Omega)+\cdots\,, \label{eqapp_au}\\
A_r&=\frac{1}{r} A_r^{(1)}(u,\Omega)+\frac{\log\frac{|u|}{2r}}{r^2}C_r^{(2)}(u,\Omega)+\frac{1}{r^2}A_r^{(2)}(u,\Omega)+\cdots\,,\label{eqapp_ar}\\
A_B&=A_B^{(0)}(u,\Omega)+\frac{\log\frac{|u|}{2r}}{r}C_B^{(1)}(u,\Omega)+\frac{1}{r} A_B^{(1)}(u,\Omega)+\cdots\,. \label{eqapp_ab}
\end{align}
Inserting them into the Lorenz gauge condition
\begin{align}
-\partial_u A_r + \partial_r (-A_u+A_r)+\frac{2}{r}(-A_u+A_r) + \frac{1}{r^2}\nabla^B A_B =0,
\end{align} 
we find  
\begin{align}
\label{p_u A_r}
\partial_u A_r^{(1)}=0\,,\quad C_u^{(1)}+\partial_u C_r^{(2)}=0\,, \quad 
-\partial_u A_r^{(2)} -\frac{1}{u} C_r^{(2)} +C_u^{(1)}- A_u^{(1)} + A_r^{(1)} +\nabla^B A_B^{(0)}=0\,,
\end{align}
where $\nabla_B$ is the covariant derivative associated with the two-sphere metric $\gamma_{AB}$, and $\nabla^B = \gamma^{BA}\nabla_A$. 

Eq.~\eqref{q+logpr} is obtained by expanding $F^{rB}$ and $F^{ru}$ in \eqref{def_Q+} with \eqref{eqapp_au}, \eqref{eqapp_ar} and \eqref{eqapp_ab}. 
Using the above expansion, $F^{rB}$ and $F^{ru}$ are computed as
\begin{align}
F^{rB}&=-\partial_u A^B +\partial_r A^B +\frac{\gamma^{BC}}{r^2}\partial_C(A_u-A_r) \nn
&=- \frac{1}{r^2} \gamma^{BC}\partial_u A^{(0)}_C-\frac{\log\frac{|u|}{2r} }{r^3} \gamma^{BA} \left(\partial_u C_A^{(1)} - \partial_A C_u^{(1)} \right)+ \mathcal{O}(r^{-3})\,,\\
F^{ru}&= \partial_u A_r -\partial_r A_u = \frac{1}{r^2} \left(A_r^{(1)} + \nabla^B A^{(0)}_B +2 C_u^{(1)} \right) + \mathcal{O}(r^{-2- \varepsilon})\,,
\end{align} 
and  the expansions lead to \eqref{q+logpr}.

The free equations of motion $\Box A_\mu=0$ in the Lorenz gauge can be written in the retarded coordinates as 
\begin{align}
&\left[\partial_r^2 -2 \partial_u \partial_r +\frac{2}{r}(-\partial_u+\partial_r)+\frac{1}{r^2}\Delta_{\mathrm{S}^2} \right]A_u=0\,,\\
&\left[\partial_r^2 -2 \partial_u \partial_r +\frac{2}{r}(-\partial_u+\partial_r)+\frac{1}{r^2}\Delta_{\mathrm{S}^2} \right]A_r-\frac{2}{r^2}(-A_u+A_r)-\frac{2}{r^3}\nabla^B A_B=0\,, \\ 
&\left[\partial_r^2 -2 \partial_u \partial_r \right]A_B+\frac{1}{r^2}\Delta_{\mathrm{S}^2}A_B +\frac{2}{r}\partial_B (-A_u+A_r)=0\,.
\end{align}
Inserting the expansions \eqref{eqapp_au}, \eqref{eqapp_ar} and \eqref{eqapp_ab}, we obtain 
\begin{align}
&\partial_u C_u^{(1)}=0, \, \,\,\,
\partial_u C_u^{(2)}=-\frac12 \Delta_{\mathrm{S}^2}C_u^{(1)}, \,\,\,\,
\partial_u A_u^{(2)} +\frac{1}{u} C_u^{(2)}= -\frac12 C_u^{(1)}+\frac12 \Delta_{\mathrm{S}^2} \left( C_u^{(1)}- A_u^{(1)}\right),\\
&\partial_u C_r^{(2)}=-C_u^{(1)}\,,\quad 
\partial_u A_r^{(2)} +\frac{1}{u} C_r^{(2)}=C_u^{(1)}-A_u^{(1)}+A_r^{(1)} -\frac12 \Delta_{\mathrm{S}^2} A_r^{(1)}+\nabla^B A_B^{(0)}\,,\\
&\partial_u C_B^{(1)}=\partial_B C_u^{(1)}\,, \quad 
\partial_u A_B^{(1)} +\frac{1}{u} C_B^{(1)}=-\partial_B (C_u^{(1)}-A_u^{(1)}+A_r^{(1)}) -\frac12 \Delta_{\mathrm{S}^2} A_B^{(0)}\,. 
\end{align}
Using the condition \eqref{p_u A_r}, we find that $A_r^{(1)}$ is a constant.

\section{Asymptotic behaviors of the massive particles}\label{app:asympt_hard}
In this appendix, we summarize our notation used in the analysis of massive particles, and  provide the concrete expressions of the matter current of a massive scalar in the asymptotic regions. 

A free massive complex scalar $\phi(x)$ can be expressed as 
\begin{align}
\phi(x)&=\int \frac{d^3p}{(2\pi)^3 2E_p} \bl b(\vec{p}) e^{ipx} +d^{\dagger}(\vec{p}) e^{-ipx}\br\,,
\end{align}
where $b(\vec{p})$ and $d(\vec{p})$ are the annihilation operators for particles and antiparticles, respectively.
The nonzero commutation relations of the creation and annihilation operators are given by
\begin{align}
[b(\vec{p}), b^\dagger(\vec{p}^{\,\prime})] =[d(\vec{p}), d^\dagger(\vec{p}^{\,\prime})] = (2\pi)^3 (2 E_p) \delta^{(3)}(\vec{p}-\vec{p}^{\,\prime})\,.
\end{align}

All massive particles go to the future timelike infinity $i^+$ not the null infinity in the asymptotic future time. 
When we work around the timelike infinity, it is convenient to introduce the following rescaled time and radial coordinates \cite{Campiglia:2015qka}:
\ba{
	\tau^{2}=t^{2}-r^{2}\ ,\ \rho=\frac{r}{\sqrt{t^{2}-r^{2}}}.
}
The Minkowski line element then takes the form
\ba{
	ds^{2}=-d\tau^{2}+\tau^{2}\, h_{\alpha\beta} d \sigma^\alpha d \sigma^\beta, 
}
where $\sigma^\alpha=(\rho, \Omega^A)$ are coordinates of the unit three-dimensional hyperbolic space $\mathbb{H}^3$ with the line element 
\begin{align}
h_{\alpha\beta} d \sigma^\alpha d \sigma^\beta=\frac{d\rho^{2}}{1+\rho^{2}}+\rho^{2}\gamma_{AB}d\Omega^{A}d\Omega^{B} .
\end{align}

In the large $\tau$ limit ($\tau \to +\infty$), using the saddle point approximation \cite{Campiglia:2015qka}, the scalar field can be expressed as 
\ba{
	\phi(\tau,\rho,\Omega)=\frac{\sqrt{m}}{2(2\pi\tau)^{3/2}}\bl b(\vec{p}) e^{-i m \tau -3\pi i/4} +d^{\dagger}(\vec{p}) e^{i m \tau +3\pi i/4}\br|_{\vec{p}=m\rho \hat{x}(\Omega)} +\mathcal{O}(\tau^{-\frac{3}{2}-\varepsilon}).
}
Therefore, in the asymptotic region, $\phi(\tau,\rho,\Omega)$ only creates (or annihilates) the (anti-)particle with localized momentum, 
\ba{\label{saddlemomentum}
\vec{p}=m\rho\hat{x}(\Omega)\ ,\ E_{p}=m\sqrt{1+\rho^{2}}
}
at the leading order. 

Then, if we ignore the interaction near the timelike infinity, the global U(1) current of the massive charged scalar with the normal ordering is given by
\ba{
  j^{mat}_{\mu}(\tau,\rho,\Omega)&=ie:\big{(}\partial_{\mu}\bar{\phi}(x)\phi(x)-\bar{\phi}(x)\partial_{\mu}\phi(x)\big{)} :\\
  &= \frac{j_{\mu}^{(3)}(\tau,\rho,\Omega)}{\tau^3}+\mathcal{O}(\tau^{-3-\varepsilon}),
}
where
\ba{
&j_{\tau}^{(3)}(\tau,\rho,\Omega)=j_{\tau}^{(3)}(\sigma)=-\frac{e m^2}{2(2\pi)^{3}}\bl b^{\dagger} b-d^{\dagger} d \br,\\
&j_{\rho}^{(3)}(\tau,\rho,\Omega)=\frac{iem}{4(2\pi)^{3}}\big{[} \bl \partial_{\rho}b^{\dagger}\,b -b^{\dagger}\partial_{\rho} b\br+i\bl b^{\dagger}\partial_{\rho}d^{\dagger}e^{2im\tau}-\partial_{\rho}b\, d e^{-2im\tau}\br-(b\leftrightarrow d)\big{]},\\
&j_{A}^{(3)}(\tau,\rho,\Omega)=\frac{iem}{4(2\pi)^{3}}\big{[} \bl \partial_A b^\dagger \, b -b^{\dagger}\partial_{A} b\br+i\bl b^{\dagger}\partial_{A}d^{\dagger}e^{2im\tau}-\partial_{A}b\, d e^{-2im\tau}\br-(b\leftrightarrow d)\big{]}.
}
Here, we have represented $b=b(m\rho\hat{x}(\Omega))$, $d=d(m\rho\hat{x}(\Omega))$ for brevity. Then one can extract the diagonal parts from the $j_{\rho}^{(3)}$ and $j_{A}^{(3)}$ by multiplying the projection operator $\frac{1}{4m^2}(\partial_{\tau}^{2}+4m^{2})$,
\ba{
&\partial_{\rho}b^{\dagger}b-b^{\dagger}\partial_{\rho} b-\partial_{\rho}d^{\dagger}d+d^{\dagger}\partial_{\rho} d
=\frac{-i(2\pi)^{3}}{em^{3}}(\partial_{\tau}^{2}+4m^{2})j_{\rho}^{(3)},\label{bd}\\
&\partial_{A}b^{\dagger}b-b^{\dagger}\partial_{A} b-\partial_{A}d^{\dagger}d+d^{\dagger}\partial_{A} d
=\frac{-i(2\pi)^{3}}{e m^{3}}(\partial_{\tau}^{2}+4m^{2})j_{A}^{(3)}.
}
Since $\frac{1}{4m^2}(\partial_{\tau}^{2}+4m^2)j_{\rho}^{(3)}$ and $\frac{1}{4m^2}(\partial_{\tau}^{2}+4m^2)j_{A}^{(3)}$ are independent of $\tau$, we represent them by $I_{\alpha}^{mat}(\sigma)$ as 
\begin{align}
I_{\alpha}^{mat}(\sigma)&\equiv \lim_{\tau\to \infty}\left[\frac{1}{4m^2}\partial_{\tau}^{2}+1\right] \tau^3 j_{\alpha}^{mat}(\tau,\sigma)\\
&=\frac{iem}{4(2\pi)^{3}} \left[\partial_{\alpha}b^{\dagger}b-b^{\dagger}\partial_{\alpha} b-\partial_{\alpha}d^{\dagger}d+d^{\dagger}\partial_{\alpha} d
\right]\,.
\end{align}

\section{Derivation of  eq.~\eqref{eq:sub_hard_h3}}\label{cal}

In this appendix, we explain some details of computation to derive \eqref{eq:sub_hard_h3}. 
Inserting \eqref{eq:QB_current} into \eqref{eq:sub_hard_1}, $Q_H^{\text{sub},+}$ is written as a sum of two parts $Q^{\text{sub},(\rho)}_H$ and $Q^{\text{sub},(\varphi)}_H$: 
\ba{Q_H^{\text{sub},+} &= Q^{\text{sub},(\rho)}_H + Q^{\text{sub},(\varphi)}_H\\
	Q^{\text{sub},(\rho)}_H&\equiv-\frac{1}{8\pi} \int d^2 \Omega \sqrt{\gamma} \int_{\mathbb{H}^3}\!\! d^3 \tilde{\sigma} \sqrt{\tilde{h}} \, \Delta_{\mathrm{S}^2} \epsilon^{(0)} \,
		I_{\rho}^{mat}(\rho, \tilde{\Omega}) \nabla^B \left[\frac{\sqrt{1+\rho^2}\, \partial_B \hat{x}(\Omega) \cdot \hat{y}(\tilde{\Omega})}{q\cdot Y} \right]
	\label{Qrho},\\
	Q^{\text{sub},(\varphi)}_H&\equiv-\frac{1}{8\pi} \int d^2 \Omega \sqrt{\gamma} \int_{\mathbb{H}^3}\!\! d^3 \tilde{\sigma} \sqrt{\tilde{h}} \, \Delta_{\mathrm{S}^2} \epsilon^{(0)} \,
	\tilde{\gamma}^{CD} I_{D}^{mat}(\rho, \tilde{\Omega})\nn
	&\quad \times \nabla^B \left[\frac{1}{q\cdot Y}
	\left\{\frac{\sqrt{1+\rho^2}}{\rho}\partial_B \hat{x} \cdot \tilde{\partial}_C\hat{y}
	-(\hat{x}\cdot\hat{y})(\partial_B \hat{x} \cdot \tilde{\partial}_C\hat{y})
	+(\hat{x} \cdot \tilde{\partial}_C\hat{y}) (\partial_B \hat{x} \cdot \hat{y})
	\right\} \right],\label{Qvarphi}
}
where $d^3 \tilde{\sigma} \sqrt{\tilde{h}} = d\rho d^2 \tilde{\Omega} \frac{\rho^2}{\sqrt{1+\rho^2}}\sqrt{\tilde{\gamma}}$. 
We now show that 
\begin{align}
Q^{\text{sub},(\rho)}_H&=\frac12\int_{\mathbb{H}^3}\!\! d^3 \sigma \sqrt{h} \frac{\sqrt{1+\rho^2}}{\rho} \left[
\rho^2 h^{\rho\rho}(\nabla^{(h)}_\rho \nabla^{(h)}_\rho  \epsilon_{\mathbb{H}^3})I_{\rho}^{mat}
+2 \rho h^{\rho\rho}(\nabla^{(h)}_\rho \epsilon_{\mathbb{H}^3})I_{\rho}^{mat}
\right],\label{Qrho_I}\\
Q^{\text{sub},(\varphi)}_H&=\frac12\int_{\mathbb{H}^3}\!\! d^3 \sigma \sqrt{h} \frac{\sqrt{1+\rho^2}}{\rho} \left[
\rho^2 h^{AB}(\nabla^{(h)}_A \nabla^{(h)}_\rho  \epsilon_{\mathbb{H}^3})I_{B}^{mat}
+2 \rho h^{AB}(\nabla^{(h)}_A \epsilon_{\mathbb{H}^3})I_{B}^{mat}
\right],\label{Qvarphi_I}
\end{align}
where $\nabla^{(h)}_\alpha$ denotes the covariant derivative compatible with the metric $h_{\alpha\beta}$ on $\mathbb{H}^3$. 
If these \eqref{Qrho_I} and \eqref{Qvarphi_I} are obtained,  eq.~\eqref{eq:sub_hard_h3} is obvious. 

In the following calculations, the formulae 
\ba{
	\partial_{A}\hat{x}\cdot \partial_{B}\hat{x}=\gamma_{AB}, \quad 
	\gamma^{AB}\partial_{A}\hat{x}_{i}\,\partial_{B}\hat{x}_{j}=\delta_{ij}-\hat{x}_{i}\,\hat{x}_{j},\quad 
	\Delta_{\mathrm{S}^2}\hat{x}_i=-2\hat{x}_{i}
}
are useful.

We first derive eq.~\eqref{Qrho_I}. 
The key equation is 
\ba{\label{softgreen}
\nabla^B \left[\frac{\partial_B \hat{x}(\Omega) \cdot \hat{y}(\tilde{\Omega})}{q(\Omega)\cdot Y(\rho,\tilde{\Omega})}\right]=\frac{4\pi}{\rho} G_{\mathbb{H}^3}(\rho, \tilde{\Omega};\Omega)-\frac{1}{\rho} \,,
}
where $G_{\mathbb{H}^3}(\rho, \tilde{\Omega};\Omega)$ was defined by eq.~\eqref{def_G3}. Furthermore, $G_{\mathbb{H}^3}(\rho, \tilde{\Omega};\Omega)$ satisfies the following property
\ba{\label{LapG}
	\Delta_{S^{2}}G_{\mathbb{H}^3}(\rho, \tilde{\Omega};\Omega)=\tilde{\Delta}_{S^{2}}G_{\mathbb{H}^3}(\rho, \tilde{\Omega};\Omega),
}
since $G_{\mathbb{H}^3}(\rho, \tilde{\Omega};\Omega)$ depends on angle $\Omega^{A}$ only through the inner product $\hat{x}(\Omega)\cdot\hat{y}(\tilde{\Omega})$. 
We thus have
\begin{align}
\int d^2 \Omega \sqrt{\gamma} \, [\Delta_{\mathrm{S}^2} \epsilon^{(0)}(\Omega)] \,
G_{\mathbb{H}^3}(\rho, \tilde{\Omega};\Omega)=\tilde{\Delta}_{\mathrm{S}^2}\int d^2 \Omega \sqrt{\gamma} \, \epsilon^{(0)}(\Omega) \,
G_{\mathbb{H}^3}(\rho, \tilde{\Omega};\Omega)
=\tilde{\Delta}_{\mathrm{S}^2}\epsilon_{\mathbb{H}^3}(\rho, \tilde{\Omega}),
\label{eq:c_1}
\end{align}
where $\epsilon_{\mathbb{H}^3}$ was defined by \eqref{def_e3}. 
By virtue of above equations, $Q^{\text{sub},(\rho)}_H$ can be written as
\ba{
	Q^{\text{sub},(\rho)}_H&=-\frac{1}{2} \int_{\mathbb{H}^3}\!\! d^3 \tilde{\sigma} \sqrt{\tilde{h}} \frac{\sqrt{1+\rho^2}}{\rho} I_{\rho}^{mat}(\rho, \tilde{\Omega}) 
	\int d^2 \Omega \sqrt{\gamma} \, \Delta_{\mathrm{S}^2} \epsilon^{(0)} \,
	 \left[G_{\mathbb{H}^3}(\rho, \tilde{\Omega};\Omega)-\frac{1}{4\pi}\right]\\
	&=-\frac{1}{2} \int_{\mathbb{H}^3}\!\! d^3 \tilde{\sigma} \sqrt{\tilde{h}} \frac{\sqrt{1+\rho^2}}{\rho} I_{\rho}^{mat}(\rho, \tilde{\Omega})\tilde{\Delta}_{\mathrm{S}^2}\epsilon_{\mathbb{H}^3}(\rho, \tilde{\Omega})\,.
}
From the first line to the second line, we have used \eqref{eq:c_1} and $\int d^2 \Omega \sqrt{\gamma}\,\Delta_{\mathrm{S}^2} \epsilon^{(0)}=0$.
In addition, since $\epsilon_{\mathbb{H}^3}(\sigma)$ is a solution of the Laplace equation on $\mathbb{H}^3$ as $\Delta_{\mathbb{H}^3}\epsilon_{\mathbb{H}^3}(\sigma)=0$,
it satisfies 
\begin{align}
\Delta_{\mathrm{S}^2}\epsilon_{\mathbb{H}^3} =-(1+\rho^2)\rho^2 \nabla^{(h)}_\rho \nabla^{(h)}_\rho  \epsilon_{\mathbb{H}^3}-2(1+\rho^2)\rho\nabla^{(h)}_\rho  \epsilon_{\mathbb{H}^3}.
\end{align}
Using this equation and noting that $h^{\rho\rho}=1+\rho^2$, eq.~\eqref{Qrho_I} can be obtained.

We next consider eq.~\eqref{Qvarphi_I}. 
In \eqref{Qvarphi}, performing a partial integration, one encounters the following quantity: 
\begin{align}
\Delta_{\mathrm{S}^2} 
\nabla^A \left[\frac{1}{q\cdot Y}
\left\{\frac{\sqrt{1+\rho^2}}{\rho}\partial_A \hat{x} \cdot \tilde{\partial}_C\hat{y}
-(\hat{x}\cdot\hat{y})(\partial_A \hat{x} \cdot \tilde{\partial}_C\hat{y})
+(\hat{x} \cdot \tilde{\partial}_C\hat{y}) (\partial_A \hat{x} \cdot \hat{y})
\right\} \right].
\label{cal1}
\end{align}
Performing the derivative, it becomes 
\begin{align}
\eqref{cal1}=\Delta_{\mathrm{S}^2} 
\left[\frac{\hat{x} \cdot \tilde{\partial}_C\hat{y}}{(q\cdot Y)^2}
\left(\frac{2}{\rho}+\rho -\sqrt{1+\rho^2}\hat{x} \cdot \hat{y}
\right)
\right].\label{cal2}
\end{align}
Performing the Laplacian, it further becomes
\ba{
	\eqref{cal2}&=-\frac{2\hat{x} \cdot \tilde{\partial}_C\hat{y}}{(q\cdot Y)^4}
	\left(\frac{2}{\rho}-\rho +\sqrt{1+\rho^2}\hat{x} \cdot \hat{y}
	\right)\nn
	&=-4\pi \frac{\sqrt{1+\rho^2}}{\rho} \left[
	\nabla^{(h)}_\rho \tilde{\nabla}^{(h)}_C G_{\mathbb{H}^3}(\rho, \tilde{\Omega};\Omega)
	+\frac{2}{\rho}\tilde{\nabla}^{(h)}_C G_{\mathbb{H}^3}(\rho, \tilde{\Omega};\Omega)
	\right]
	\label{cal3}.
}
Therefore, \eqref{Qvarphi} can be written as 
\begin{align}
Q^{\text{sub},(\varphi)}_H&=\frac{1}{2} \int_{\mathbb{H}^3}\!\! d^3 \tilde{\sigma} \sqrt{\tilde{h}} \frac{\sqrt{1+\rho^2}}{\rho}\tilde{\gamma}^{CD} I_{D}^{mat}(\rho, \tilde{\Omega})\nn
&\qquad\times \int d^2 \Omega \sqrt{\gamma}  \epsilon^{(0)} \left[
\nabla^{(h)}_\rho \tilde{\nabla}^{(h)}_C G_{\mathbb{H}^3}(\rho, \tilde{\Omega};\Omega)
+\frac{2}{\rho}\tilde{\nabla}^{(h)}_C G_{\mathbb{H}^3}(\rho, \tilde{\Omega};\Omega)
\right]\nn
&=\frac{1}{2} \int_{\mathbb{H}^3}\!\! d^3 \tilde{\sigma} \sqrt{\tilde{h}} \frac{\sqrt{1+\rho^2}}{\rho}\tilde{\gamma}^{CD} I_{D}^{mat}(\rho, \tilde{\Omega})
\left[
\nabla^{(h)}_\rho \tilde{\nabla}^{(h)}_C \epsilon_{\mathbb{H}^3}(\rho, \tilde{\Omega})
+\frac{2}{\rho}\tilde{\nabla}^{(h)}_C \epsilon_{\mathbb{H}^3}(\rho, \tilde{\Omega})
\right]\nn
&=\frac{1}{2} \int_{\mathbb{H}^3}\!\! d^3 \sigma \sqrt{h} \frac{\sqrt{1+\rho^2}}{\rho}h^{CD} I_{D}^{mat}(\rho, \Omega)
\left[\rho^2\nabla^{(h)}_\rho \nabla^{(h)}_C \epsilon_{\mathbb{H}^3}(\rho, \Omega)
+2\rho\nabla^{(h)}_C \epsilon_{\mathbb{H}^3}(\rho, \Omega)
\right],
\end{align}
where we have renamed the integration variables and used $\gamma^{AB}= \rho^2 h^{AB}$ in the last line. 
Thus, we have obtained eq.~\eqref{Qvarphi_I}.

\bibliographystyle{utphys}

\bibliography{Ref}

\providecommand{\href}[2]{#2}\begingroup\raggedright\begin{thebibliography}{10}

\bibitem{Strominger:2017zoo}
A.~Strominger, ``{Lectures on the Infrared Structure of Gravity and Gauge
  Theory},''
\href{http://arxiv.org/abs/1703.05448}{{\ttfamily arXiv:1703.05448 [hep-th]}}.

\bibitem{Campiglia:2017dpg}
M.~Campiglia, L.~Coito, and S.~Mizera, ``{Can scalars have asymptotic
  symmetries?},'' \href{http://dx.doi.org/10.1103/PhysRevD.97.046002}{{\em
  Phys. Rev.} {\bfseries D97} no.~4, (2018) 046002},
\href{http://arxiv.org/abs/1703.07885}{{\ttfamily arXiv:1703.07885 [hep-th]}}.

\bibitem{Hamada:2017atr}
Y.~Hamada and S.~Sugishita, ``{Soft pion theorem, asymptotic symmetry and new
  memory effect},'' \href{http://dx.doi.org/10.1007/JHEP11(2017)203}{{\em JHEP}
  {\bfseries 11} (2017) 203},
\href{http://arxiv.org/abs/1709.05018}{{\ttfamily arXiv:1709.05018 [hep-th]}}.

\bibitem{Campiglia:2017xkp}
M.~Campiglia and L.~Coito, ``{Asymptotic charges from soft scalars in even
  dimensions},'' \href{http://dx.doi.org/10.1103/PhysRevD.97.066009}{{\em Phys.
  Rev.} {\bfseries D97} no.~6, (2018) 066009},
\href{http://arxiv.org/abs/1711.05773}{{\ttfamily arXiv:1711.05773 [hep-th]}}.

\bibitem{Yennie:1961ad}
D.~R. Yennie, S.~C. Frautschi, and H.~Suura, ``{The infrared divergence
  phenomena and high-energy processes},''
\href{http://dx.doi.org/10.1016/0003-4916(61)90151-8}{{\em Annals Phys.}
  {\bfseries 13} (1961) 379--452}.

\bibitem{Weinberg:1965nx}
S.~Weinberg, ``{Infrared photons and gravitons},''
\href{http://dx.doi.org/10.1103/PhysRev.140.B516}{{\em Phys. Rev.} {\bfseries
  140} (1965) B516--B524}.

\bibitem{He:2014cra}
T.~He, P.~Mitra, A.~P. Porfyriadis, and A.~Strominger, ``{New Symmetries of
  Massless QED},'' \href{http://dx.doi.org/10.1007/JHEP10(2014)112}{{\em JHEP}
  {\bfseries 10} (2014) 112},
\href{http://arxiv.org/abs/1407.3789}{{\ttfamily arXiv:1407.3789 [hep-th]}}.

\bibitem{Campiglia:2015qka}
M.~Campiglia and A.~Laddha, ``{Asymptotic symmetries of QED and Weinberg’s
  soft photon theorem},'' \href{http://dx.doi.org/10.1007/JHEP07(2015)115}{{\em
  JHEP} {\bfseries 07} (2015) 115},
\href{http://arxiv.org/abs/1505.05346}{{\ttfamily arXiv:1505.05346 [hep-th]}}.

\bibitem{Kapec:2015ena}
D.~Kapec, M.~Pate, and A.~Strominger, ``{New Symmetries of QED},''
  \href{http://dx.doi.org/10.4310/ATMP.2017.v21.n7.a7}{{\em Adv. Theor. Math.
  Phys.} {\bfseries 21} (2017) 1769--1785},
\href{http://arxiv.org/abs/1506.02906}{{\ttfamily arXiv:1506.02906 [hep-th]}}.

\bibitem{Low:1954kd}
F.~E. Low, ``{Scattering of light of very low frequency by systems of spin
  1/2},''
\href{http://dx.doi.org/10.1103/PhysRev.96.1428}{{\em Phys. Rev.} {\bfseries
  96} (1954) 1428--1432}.

\bibitem{Low:1958sn}
F.~E. Low, ``{Bremsstrahlung of very low-energy quanta in elementary particle
  collisions},''
\href{http://dx.doi.org/10.1103/PhysRev.110.974}{{\em Phys. Rev.} {\bfseries
  110} (1958) 974--977}.

\bibitem{Burnett:1967km}
T.~H. Burnett and N.~M. Kroll, ``{Extension of the low soft photon theorem},''
\href{http://dx.doi.org/10.1103/PhysRevLett.20.86}{{\em Phys. Rev. Lett.}
  {\bfseries 20} (1968) 86}.

\bibitem{GellMann:1954kc}
M.~Gell-Mann and M.~L. Goldberger, ``{Scattering of low-energy photons by
  particles of spin 1/2},''
\href{http://dx.doi.org/10.1103/PhysRev.96.1433}{{\em Phys. Rev.} {\bfseries
  96} (1954) 1433--1438}.

\bibitem{Lysov:2014csa}
V.~Lysov, S.~Pasterski, and A.~Strominger, ``{Low’s Subleading Soft Theorem
  as a Symmetry of QED},''
  \href{http://dx.doi.org/10.1103/PhysRevLett.113.111601}{{\em Phys. Rev.
  Lett.} {\bfseries 113} no.~11, (2014) 111601},
\href{http://arxiv.org/abs/1407.3814}{{\ttfamily arXiv:1407.3814 [hep-th]}}.

\bibitem{Campiglia:2016hvg}
M.~Campiglia and A.~Laddha, ``{Subleading soft photons and large gauge
  transformations},'' \href{http://dx.doi.org/10.1007/JHEP11(2016)012}{{\em
  JHEP} {\bfseries 11} (2016) 012},
\href{http://arxiv.org/abs/1605.09677}{{\ttfamily arXiv:1605.09677 [hep-th]}}.

\bibitem{Conde:2016csj}
E.~Conde and P.~Mao, ``{Remarks on asymptotic symmetries and the subleading
  soft photon theorem},''
  \href{http://dx.doi.org/10.1103/PhysRevD.95.021701}{{\em Phys. Rev.}
  {\bfseries D95} no.~2, (2017) 021701},
\href{http://arxiv.org/abs/1605.09731}{{\ttfamily arXiv:1605.09731 [hep-th]}}.

\bibitem{ZelPol}
Y.~B. Zel'dovich and A.~G. Polnarev, ``{Radiation of gravitational waves by a
  cluster of superdense stars},'' {\em Soviet Astronomy} {\bfseries 51} (1974)
  .

\bibitem{Braginsky1987}
V.~B. Braginsky and K.~S. Thorne, ``{Gravitational-wave bursts with memory and
  experimental prospects},''
{\em Nature} {\bfseries 327} (1987) .

\bibitem{Christodoulou:1991cr}
D.~Christodoulou, ``{Nonlinear nature of gravitation and gravitational wave
  experiments},''
\href{http://dx.doi.org/10.1103/PhysRevLett.67.1486}{{\em Phys. Rev. Lett.}
  {\bfseries 67} (1991) 1486--1489}.

\bibitem{Thorne:1992sdb}
K.~S. Thorne, ``{Gravitational-wave bursts with memory: The Christodoulou
  effect},''
\href{http://dx.doi.org/10.1103/PhysRevD.45.520}{{\em Phys. Rev.} {\bfseries
  D45} no.~2, (1992) 520--524}.

\bibitem{Bieri:2013hqa}
L.~Bieri and D.~Garfinkle, ``{An electromagnetic analogue of gravitational wave
  memory},'' \href{http://dx.doi.org/10.1088/0264-9381/30/19/195009}{{\em
  Class. Quant. Grav.} {\bfseries 30} (2013) 195009},
\href{http://arxiv.org/abs/1307.5098}{{\ttfamily arXiv:1307.5098 [gr-qc]}}.

\bibitem{Tolish:2014bka}
A.~Tolish and R.~M. Wald, ``{Retarded Fields of Null Particles and the Memory
  Effect},'' \href{http://dx.doi.org/10.1103/PhysRevD.89.064008}{{\em Phys.
  Rev.} {\bfseries D89} no.~6, (2014) 064008},
\href{http://arxiv.org/abs/1401.5831}{{\ttfamily arXiv:1401.5831 [gr-qc]}}.

\bibitem{Susskind:2015hpa}
L.~Susskind, ``{Electromagnetic Memory},''
\href{http://arxiv.org/abs/1507.02584}{{\ttfamily arXiv:1507.02584 [hep-th]}}.

\bibitem{Strominger:2014pwa}
A.~Strominger and A.~Zhiboedov, ``{Gravitational Memory, BMS Supertranslations
  and Soft Theorems},'' \href{http://dx.doi.org/10.1007/JHEP01(2016)086}{{\em
  JHEP} {\bfseries 01} (2016) 086},
\href{http://arxiv.org/abs/1411.5745}{{\ttfamily arXiv:1411.5745 [hep-th]}}.

\bibitem{Pasterski:2015tva}
S.~Pasterski, A.~Strominger, and A.~Zhiboedov, ``{New Gravitational
  Memories},'' \href{http://dx.doi.org/10.1007/JHEP12(2016)053}{{\em JHEP}
  {\bfseries 12} (2016) 053},
\href{http://arxiv.org/abs/1502.06120}{{\ttfamily arXiv:1502.06120 [hep-th]}}.

\bibitem{Mao:2017wvx}
P.~Mao and H.~Ouyang, ``{Note on soft theorems and memories in even
  dimensions},'' \href{http://dx.doi.org/10.1016/j.physletb.2017.08.064}{{\em
  Phys. Lett.} {\bfseries B774} (2017) 715--722},
\href{http://arxiv.org/abs/1707.07118}{{\ttfamily arXiv:1707.07118 [hep-th]}}.

\bibitem{Hamada:2018cjj}
Y.~Hamada and S.~Sugishita, ``{Notes on the gravitational, electromagnetic and
  axion memory effects},''
\href{http://arxiv.org/abs/1803.00738}{{\ttfamily arXiv:1803.00738 [hep-th]}}.

\bibitem{Mao:2017axa}
P.~Mao, H.~Ouyang, J.-B. Wu, and X.~Wu, ``{New electromagnetic memories and
  soft photon theorems},''
  \href{http://dx.doi.org/10.1103/PhysRevD.95.125011}{{\em Phys. Rev.}
  {\bfseries D95} no.~12, (2017) 125011},
\href{http://arxiv.org/abs/1703.06588}{{\ttfamily arXiv:1703.06588 [hep-th]}}.

\bibitem{He:2014laa}
T.~He, V.~Lysov, P.~Mitra, and A.~Strominger, ``{BMS supertranslations and
  Weinberg’s soft graviton theorem},''
  \href{http://dx.doi.org/10.1007/JHEP05(2015)151}{{\em JHEP} {\bfseries 05}
  (2015) 151},
\href{http://arxiv.org/abs/1401.7026}{{\ttfamily arXiv:1401.7026 [hep-th]}}.

\bibitem{Ashtekar:1981sf}
A.~Ashtekar, ``{Asymptotic Quantization of the Gravitational Field},''
\href{http://dx.doi.org/10.1103/PhysRevLett.46.573}{{\em Phys. Rev. Lett.}
  {\bfseries 46} (1981) 573--576}.

\bibitem{Frolov:1977bp}
V.~P. Frolov, ``{Null Surface Quantization and Quantum Field Theory in
  Asymptotically Flat Space-Time},''
\href{http://dx.doi.org/10.1002/prop.19780260902}{{\em Fortsch. Phys.}
  {\bfseries 26} (1978) 455}.

\bibitem{Becchi:1975nq}
C.~Becchi, A.~Rouet, and R.~Stora, ``{Renormalization of Gauge Theories},''
\href{http://dx.doi.org/10.1016/0003-4916(76)90156-1}{{\em Annals Phys.}
  {\bfseries 98} (1976) 287--321}.

\bibitem{Tyutin:1975qk}
I.~V. Tyutin, ``{Gauge Invariance in Field Theory and Statistical Physics in
  Operator Formalism},''
\href{http://arxiv.org/abs/0812.0580}{{\ttfamily arXiv:0812.0580 [hep-th]}}.

\bibitem{Hamada:2017bgi}
Y.~Hamada, M.-S. Seo, and G.~Shiu, ``{Electromagnetic Duality and the Electric
  Memory Effect},'' \href{http://dx.doi.org/10.1007/JHEP02(2018)046}{{\em JHEP}
  {\bfseries 02} (2018) 046},
\href{http://arxiv.org/abs/1711.09968}{{\ttfamily arXiv:1711.09968 [hep-th]}}.

\bibitem{Hamada:2018vrw}
Y.~Hamada and G.~Shiu, ``{An infinite set of soft theorems in gauge/gravity
  theories as Ward-Takahashi identities},''
\href{http://arxiv.org/abs/1801.05528}{{\ttfamily arXiv:1801.05528 [hep-th]}}.

\bibitem{Kugo:1981hm}
T.~Kugo and S.~Uehara, ``{General Procedure of Gauge Fixing Based on {BRS}
  Invariance Principle},''
\href{http://dx.doi.org/10.1016/0550-3213(82)90449-7}{{\em Nucl. Phys.}
  {\bfseries B197} (1982) 378--384}.

\bibitem{Kugo:1977zq}
T.~Kugo and I.~Ojima, ``{Manifestly Covariant Canonical Formulation of
  Yang-Mills Field Theories: Physical State Subsidiary Conditions and Physical
  S Matrix Unitarity},''
\href{http://dx.doi.org/10.1016/0370-2693(78)90765-7}{{\em Phys. Lett.}
  {\bfseries 73B} (1978) 459--462}.

\bibitem{Gupta:1949rh}
S.~N. Gupta, ``{Theory of longitudinal photons in quantum electrodynamics},''
\href{http://dx.doi.org/10.1088/0370-1298/63/7/301}{{\em Proc. Phys. Soc.}
  {\bfseries A63} (1950) 681--691}.

\bibitem{Bleuler:1950cy}
K.~Bleuler, ``{A New method of treatment of the longitudinal and scalar
  photons},''
{\em Helv. Phys. Acta} {\bfseries 23} (1950) 567--586.

\bibitem{Cachazo:2014fwa}
F.~Cachazo and A.~Strominger, ``{Evidence for a New Soft Graviton Theorem},''
\href{http://arxiv.org/abs/1404.4091}{{\ttfamily arXiv:1404.4091 [hep-th]}}.

\bibitem{Hawking:2016msc}
S.~W. Hawking, M.~J. Perry, and A.~Strominger, ``{Soft Hair on Black Holes},''
  \href{http://dx.doi.org/10.1103/PhysRevLett.116.231301}{{\em Phys. Rev.
  Lett.} {\bfseries 116} no.~23, (2016) 231301},
\href{http://arxiv.org/abs/1601.00921}{{\ttfamily arXiv:1601.00921 [hep-th]}}.

\bibitem{Chu:2018tzu}
C.-S. Chu and Y.~Koyama, ``{Soft Hair of Dynamical Black Hole and Hawking
  Radiation},'' \href{http://dx.doi.org/10.1007/JHEP04(2018)056}{{\em JHEP}
  {\bfseries 04} (2018) 056},
\href{http://arxiv.org/abs/1801.03658}{{\ttfamily arXiv:1801.03658 [hep-th]}}.

\bibitem{Seraj:2016jxi}
A.~Seraj, ``{Multipole charge conservation and implications on electromagnetic
  radiation},'' \href{http://dx.doi.org/10.1007/JHEP06(2017)080}{{\em JHEP}
  {\bfseries 06} (2017) 080},
\href{http://arxiv.org/abs/1610.02870}{{\ttfamily arXiv:1610.02870 [hep-th]}}.

\bibitem{Compere:2017wrj}
G.~Compère, R.~Oliveri, and A.~Seraj, ``{Gravitational multipole moments from
  Noether charges},''
\href{http://arxiv.org/abs/1711.08806}{{\ttfamily arXiv:1711.08806 [hep-th]}}.

\bibitem{Chung:1965zza}
V.~Chung, ``{Infrared Divergence in Quantum Electrodynamics},''
\href{http://dx.doi.org/10.1103/PhysRev.140.B1110}{{\em Phys. Rev.} {\bfseries
  140} (1965) B1110--B1122}.

\bibitem{Kulish:1970ut}
P.~P. Kulish and L.~D. Faddeev, ``{Asymptotic conditions and infrared
  divergences in quantum electrodynamics},''
  \href{http://dx.doi.org/10.1007/BF01066485}{{\em Theor. Math. Phys.}
  {\bfseries 4} (1970) 745}.
[Teor. Mat. Fiz.4,153(1970)].

\bibitem{Ware:2013zja}
J.~Ware, R.~Saotome, and R.~Akhoury, ``{Construction of an asymptotic S matrix
  for perturbative quantum gravity},''
  \href{http://dx.doi.org/10.1007/JHEP10(2013)159}{{\em JHEP} {\bfseries 10}
  (2013) 159},
\href{http://arxiv.org/abs/1308.6285}{{\ttfamily arXiv:1308.6285 [hep-th]}}.

\bibitem{Mirbabayi:2016axw}
M.~Mirbabayi and M.~Porrati, ``{Dressed Hard States and Black Hole Soft
  Hair},'' \href{http://dx.doi.org/10.1103/PhysRevLett.117.211301}{{\em Phys.
  Rev. Lett.} {\bfseries 117} no.~21, (2016) 211301},
\href{http://arxiv.org/abs/1607.03120}{{\ttfamily arXiv:1607.03120 [hep-th]}}.

\bibitem{Gabai:2016kuf}
B.~Gabai and A.~Sever, ``{Large gauge symmetries and asymptotic states in
  QED},'' \href{http://dx.doi.org/10.1007/JHEP12(2016)095}{{\em JHEP}
  {\bfseries 12} (2016) 095},
\href{http://arxiv.org/abs/1607.08599}{{\ttfamily arXiv:1607.08599 [hep-th]}}.

\bibitem{Kapec:2017tkm}
D.~Kapec, M.~Perry, A.-M. Raclariu, and A.~Strominger, ``{Infrared Divergences
  in QED, Revisited},''
  \href{http://dx.doi.org/10.1103/PhysRevD.96.085002}{{\em Phys. Rev.}
  {\bfseries D96} no.~8, (2017) 085002},
\href{http://arxiv.org/abs/1705.04311}{{\ttfamily arXiv:1705.04311 [hep-th]}}.

\bibitem{Choi:2017bna}
S.~Choi, U.~Kol, and R.~Akhoury, ``{Asymptotic Dynamics in Perturbative Quantum
  Gravity and BMS Supertranslations},''
  \href{http://dx.doi.org/10.1007/JHEP01(2018)142}{{\em JHEP} {\bfseries 01}
  (2018) 142},
\href{http://arxiv.org/abs/1708.05717}{{\ttfamily arXiv:1708.05717 [hep-th]}}.

\bibitem{Choi:2017ylo}
S.~Choi and R.~Akhoury, ``{BMS Supertranslation Symmetry Implies Faddeev-Kulish
  Amplitudes},'' \href{http://dx.doi.org/10.1007/JHEP02(2018)171}{{\em JHEP}
  {\bfseries 02} (2018) 171},
\href{http://arxiv.org/abs/1712.04551}{{\ttfamily arXiv:1712.04551 [hep-th]}}.

\bibitem{Carney:2018ygh}
D.~Carney, L.~Chaurette, D.~Neuenfeld, and G.~Semenoff, ``{On the need for soft
  dressing},''
\href{http://arxiv.org/abs/1803.02370}{{\ttfamily arXiv:1803.02370 [hep-th]}}.

\end{thebibliography}\endgroup
\end{document}